\documentclass[aps,superscriptaddress,twocolumn]{revtex4}

\usepackage{latexsym}
\usepackage{amssymb}
\usepackage{amsmath}
\usepackage{graphicx}
\usepackage{bm}
\usepackage{epsf}
\usepackage{epstopdf}
\usepackage[colorlinks,
            citecolor=blue,
            anchorcolor=red,
            menucolor=red,
            linkcolor=red,
            filecolor=red,
            runcolor=red,
            urlcolor=blue,
            frenchlinks=red]{hyperref}

\newcommand{\fm}{\mathrm{fm}}
\usepackage[usenames]{color}
\usepackage[normalem]{ulem}  %

\newcommand{\comment}[1]{}

\renewcommand\sout{\bgroup \color{red} \ULdepth=-.5ex \ULset}

\begin{document}

\newcommand{\YSU}{Key Laboratory for Microstructural Material Physics of Hebei Province, School of Science, Yanshan University, Qinhuangdao, 066004, China}
\newcommand{\QDU}{Science School, Qingdao University of Technology, Qingdao 266000, China}
\newcommand{\RCQDU}{The Research Center of Theoretical Physics, Qingdao University of Technology, Qingdao 266033, China}
\newcommand{\PeU}{School of Science, China University of Petroleum (East China), Qingdao 266580, China}

\title{The mixed phase quark core in massive hybrid stars}

\author{Xuhao Wu}~\email{wuhaobird@gmail.com}
\affiliation{\YSU}
\author{Peng-Cheng Chu}~\email{kyois@126.com}
\affiliation{\QDU}
\affiliation{\RCQDU}
\author{Min Ju}~\email{jumin@upc.edu.cn}
\affiliation{\PeU}
\author{He Liu}~\email{liuhe@qut.edu.cn}
\affiliation{\QDU}
\affiliation{\RCQDU}

\begin{abstract}
We investigate the properties of hybrid star and the mixed phase core to explore the radius ratio of the mixed phase in hybrid star.
In the context of observed massive neutron stars (NSs), we examine the internal structure, phase transitions, and the impacts of the equation of state (EOS) in maximum hybrid star.
We investigate the stiffness changes in the EOS during the hadron-quark phase transition within the hybrid stars.
The relativistic mean-field (RMF) model is used to describe hadronic matter, 
while to the represent quark matter, the Nambu-Jona-Lasinio (NJL) model is applied.
We explore the strength of vector coupling in quark matter, 
which delays the onset density of the mixed phase and reduces the size of the mixed-phase core in a hybrid star,
but does not exhibit a clear correlation with the central density.
In a hybrid star with a maximum mass of approximately 2 solar masses ($M_\odot$), 
a mixed-phase core of $\sim$5 km may exist, comprising about $40\%$ of the total radius.
However, our results do not support the existence of a sizable quark core containing the mixed phase ($R_{\rm{MP}}>1/2~R_{\rm{total}}$) 
for the maximum-mass hybrid star or for a 2~$M_\odot$ massive star.
\end{abstract}

\maketitle

\section{Introduction}
\label{sec:intro}

The equation of state (EOS) plays a crucial role in the fields of nuclear physics and astrophysics~\cite{Baym2018}, 
as it determines the composition and properties of compact stars.
It also sheds light on the nature of strong interactions within these stars. 
The quantum chromodynamics (QCD) phase transition at finite temperature and low baryon density supports the hot dense quark matter exist, 
while cold dense quark matter may only be found in neutron stars (NSs).
The question of whether quark matter can exist in NSs has been a topic of interest 
for researchers in several decades~\cite{Baym1976,Glendenning1992,Mariani2024}, 
which remains an open question. 
It is challenging to draw firm conclusions about the presence of quark matter inside NSs based solely on measurements of mass or radius. 
However, the sensitivity of core g-mode oscillations to the presence of quark matter may help mitigate this difficulty~\cite{Wei2020,Jaikumar2021}.
At high density, the Bodmer–Witten assumption that strange quark matter composed of 
$u$, $d$ and $s$ quarks and leptons in $\beta$ equilibrium condition may actually be the true ground state of matter~\cite{Witten1984}, 
exhibiting a lower energy per baryon compared to both $ud$ quark matter and nucleon matter.
Whether strange quark matter is absolutely stable is model dependent~\cite{Liu2023}, 
if this absolutely stable condition (the Bodmer–Witten assumption) is satisfied, a conversion from NS to strange quark star may occur.
Otherwise, the cold dense quark matter may found in NS, which is usually called hybrid star.
Overall, it is possible that a first-order hadron-quark phase transition occurs within the core of massive hybrid star~\cite{Glendenning2001,Weber2005,Han2019,Brandes2023}, 
where the density could reach $5\sim10$ times saturation density $n_0$.
Han et al.~\cite{Han2019} explored various methods for implementing the hadron-quark phase transition, 
specifically, first-order transitions with Maxwell construction and Gibbs construction, 
as well as smooth crossover transitions and quarkyonic matter.
The framework of hybrid star with twin star under Maxwell construction is discussed in~\cite{Li2021,Podder2024}.
Li et al.~\cite{Li2023apj} proposed a model featuring sequential phase transitions from hadronic matter to low-density and subsequently to high-density quark matter phases.

In the past decade, significant progress in astronomical observations has provided abundant data on NS mass, radius, and tidal deformability. 
These constraints have spurred explorations into theories of dense matter, narrowing down the range of plausible strong interaction theories. 
Recent Shapiro delay measurements of massive neutron star PSR J0740+6620~\cite{Cromartie2020,Fonseca2021}, 
provide a lower limit of NS maximum mass, $2.08\pm{0.07}~M_\odot$ (68.3\% credibility interval), 
similar restrictions are also imposed by PSR J1614-2230 ($1.908\pm0.016~M_\odot$)~\cite{Arzoumanian2018}
and PSR J0348+0432 ($2.01\pm0.04~M_\odot$)~\cite{Antoniadis2013}, respectively.
These results provide constraints on the EOSs,
that disallow configurations unable to support $\sim2~M_\odot$ NSs. 
Additionally, the black-widow pulsar, PSR J0952-0607 has a reported mass of $2.35\pm{0.17}~M_\odot$~\cite{Romani2022} (68.3\% credibility interval), making it the heaviest NS observed.
The inferred mass $M$ and equatorial radius $R$ from x-ray data collected by the Neutron Star Interior Composition Explorer (NICER) for PSR J0030+0451 are estimated to be $M=1.34_{-0.16}^{+0.15}~M_{\odot}$, $R=12.71_{-1.19}^{+1.14}~\mathrm{km}$~\cite{Riley2019} 
and $M=1.44_{-0.14}^{+0.15}~M_{\odot}$, $R=13.02_{-1.06}^{+1.24}~\mathrm{km}$~\cite{Miller2019} (68\% credible interval), respectively.
Moreover, observations of PSR J0740+6620 from NICER (68.3\% credibility interval) suggest a mass of 
$M=2.072_{-0.066}^{+0.067}~M_\odot$ and an equatorial radius of $R=12.39_{-0.98}^{+1.30}~\mathrm{km}$~\cite{Riley2021}. 
Additionally, another estimate indicates $M=2.08\pm0.07~M_\odot$ and $R=13.7_{-1.5}^{+2.6}~\mathrm{km}$~\cite{Miller2021}.
These massive NS observations above ruled out as unable to support NSs with masses $\sim2~M_\odot$.
In addition to these compact objects, there are
other significant estimates, including the possibility 
that the secondary component of GW190814 may be the most massive NS to date~\cite{Abbott2020}.
The GW170817 event provided constraints on tidal deformability, with $\Lambda_{1.4}=190^{+390}_{-120}$~\cite{Abbott2017,Abbott2018}.
A roughly $1.4~M_\odot$ NS radius, consistent with the tidal deformability up limit, 
refers $R_{1.4}\leq13.6~\rm{km}$~\cite{Annala2018}.
The advancements from NICER and gravitational wave detections have opened new avenues for NS exploration.
These multi-messenger astronomical observations can be used to constrain the parameters in different models~\cite{Ferreira2021,Liu2023b,Gholami2024a}.

On one hand, the mass constraints for massive NSs require a stiff EOS;
on the other hand, the radius or tidal deformability limits suggest a relatively soft EOS in the low-density range. 
To reconcile these conflicting constraints, 
numerous studies have explored the inclusion of quark degrees of freedom in NSs~\cite{Alford2019,Bauswein2019,Annala2020,Ferreira2021,Huang2022,Liu2023,Liu2023b}. 
Recently, it has been suggested that observations of gravitational waves from binary neutron star mergers indicate the possibility of a hadron-quark phase transition~\cite{Annala2020}. 
This conclusion is based on a model-independent speed-of-sound interpolation between chiral effective field theory (CEFT) and perturbative quantum chromodynamics (pQCD) EOSs. 
Their findings imply the existence of a relatively large pure quark core in massive massive NSs~\cite{Annala2020}. 
This conclusion is further supported by Refs.~\cite{Liu2023b,Ferreira2021}, which include additional quark interactions.
In these studies, the eight-quark vector interaction and the four-quark isovector-vector interaction play significantly different roles in determining the size of the quark core.

The purpose of this study is to explore the changes in the stiffness of the EOS during the transition from hadronic matter to quark matter, 
and to investigate the properties of the maximum-mass hybrid star that incorporates quark degrees of freedom, 
particularly focusing on the size of the quark matter core.
A quantitative description of EOS stiffness provides insight into the density of hybrid star matter required to satisfy observational constraints and strong interaction strengths. 
Furthermore, the presence of quark matter in the core of massive hybrid stars offers additional insights into the deconfinement phase transition.
Describing both hadronic and quark matter within a unified framework presents several challenges, 
as these phases typically involve different types of particles.
Distinct models are typically selected to describe hadronic and quark matter. 
For hadronic matter, we employ the relativistic mean-field (RMF) model. 
For quark matter, we utilize the Nambu-Jona-Lasinio (NJL) model, 
which can spontaneously describe the restoration of chiral symmetry by calculating quark condensates.
The Gibbs construction~\cite{Glendenning1992,Maruyama2008,Wu2019} is adopted for modeling the hadron-quark mixed phase.
In the Gibbs construction both hadronic matter and quark matter can coexist within a density region characterized by dynamic equilibrium, 
baryon chemical potential equilibrium and global charge neutrality. 
The transition from the hadronic phase to the quark phase leads to a softening of the EOS, 
resulting in a lower maximum mass for hybrid stars compared to their pure hadronic counterparts, due to the increase in degrees of freedom.
The model independent calculations~\cite{Annala2020} and phenomenological models with a first-order phase transition~\cite{Liu2023b,Ferreira2021} support the existence of a large quark core in massive hybrid star.
However, our results using RMF-NJL framework suggest that a Gibbs phase transition is insufficient to produce a sizable quark core 
that occupies nearly half the radius of massive hybrid stars, as indicated in Ref.~\cite{Annala2020}.
Another commonly used Maxwell construction leads to an unstable hybrid star when quark matter appears inside a neutron star using the RMF-NJL framework. 
Therefore, there is no quark core under the Maxwell construction in this work.
In this study, hyperons in hadronic matter are neglected mainly due to the ``hyperon puzzle" problem. 
Hyperons would soften the EOS and may prevent the formation of NSs with masses above 2 $M_\odot$.
This issue requires introducing additional effects, such as three-body forces, which complicate the discussion.
Additionally, a softer EOS caused by hyperons would delay the onset density of quarks and lead to a smaller quark core. 
Furthermore, the onset densities of quarks and the lightest hyperon are similar, making it difficult to distinguish the effects of hyperons from those of quark degree on the properties of the EOS of hybrid star matter.
The recent observational constraints mentioned above have been taken into account. 
Within the framework of this study, quarks typically emerge beyond the central density of a 1.4 $M_\odot$ NS, 
rendering the constraints on tidal deformability and the corresponding radius irrelevant to the quark degrees of freedom.

This article is constructed as follows.
In Section~\ref{sec:model}, we provide a concise overview of the RMF model and the NJL model. 
We also discuss the hadron-quark phase transition under Gibbs equilibrium.
In Section~\ref{sec:results}, we present the numerical results of the hadron-quark phase transition,
including the changes in the stiffness of the EOS and the size of the quark matter core.
Finally, Section~\ref{sec:summary} offers a summary of our findings.

\section{The theoretical model}
\label{sec:model}
\subsection{Hadronic matter phase}

The RMF model is employed to describe the hadronic matter, 
where nucleons interact through the exchange of isoscalar-scalar meson $\sigma$, isoscalar-vector meson $\omega$, and isovector-vector meson $\rho$.
The Lagrangian density for the hadronic matter, comprising nucleons ($p$ and $n$) and
leptons ($e$ and $\mu$) is written as
\begin{eqnarray}
\label{eq:LRMF}
\mathcal{L}_{\rm{RMF}} & = & \sum_{i=p,n}\bar{\psi}_i
\left\{i\gamma_{\mu}\partial^{\mu}-\left(M+g_{\sigma}\sigma\right) \right.\notag \\
&& \left. -\gamma_{\mu} \left[g_{\omega}\omega^{\mu} +\frac{g_{\rho}}{2}\tau_a\rho^{a\mu}
\right]\right\}\psi_i  \notag \\
&& +\frac{1}{2}\partial_{\mu}\sigma\partial^{\mu}\sigma -\frac{1}{2}%
m^2_{\sigma}\sigma^2-\frac{1}{3}g_{2}\sigma^{3} -\frac{1}{4}g_{3}\sigma^{4}
\notag \\
&& -\frac{1}{4}W_{\mu\nu}W^{\mu\nu} +\frac{1}{2}m^2_{\omega}\omega_{\mu}%
\omega^{\mu} +\frac{1}{4}c_{3}\left(\omega_{\mu}\omega^{\mu}\right)^2  \notag
\\
&& -\frac{1}{4}R^a_{\mu\nu}R^{a\mu\nu} +\frac{1}{2}m^2_{\rho}\rho^a_{\mu}%
\rho^{a\mu} \notag \\
&& +\Lambda_{\rm{v}} \left(g_{\omega}^2
\omega_{\mu}\omega^{\mu}\right)
\left(g_{\rho}^2\rho^a_{\mu}\rho^{a\mu}\right) \notag\\
&& +\sum_{l=e,\mu}\bar{\psi}_{l}
  \left( i\gamma_{\mu }\partial^{\mu }-m_{l}\right)\psi_l,
\end{eqnarray}
where $W^{\mu\nu}$ and $R^{a\mu\nu}$ denote the antisymmetric field tensors associated with $\omega^{\mu}$ and $\rho^{a\mu}$, respectively.
Within the RMF framework, meson fields are treated as classical fields, 
and the field operators are replaced with their corresponding expectation values. 
For a static system, the non-vanishing expectation values are
$\sigma =\left\langle \sigma \right\rangle$,
$\omega =\left\langle\omega^{0}\right\rangle$, and $\rho =\left\langle \rho^{30} \right\rangle$.
In uniform hadronic matter, the equations of motion for meson fields can be expressed as
\begin{eqnarray}
m_{\sigma }^{2}\sigma +g_{2}\sigma ^{2}+g_{3}\sigma^{3}
&=&-g_{\sigma }\left( n_{p}^{s}+n_{n}^{s}\right) ,
\label{eq:eqms} \\
m_{\omega }^{2}\omega +c_{3}\omega^{3}
+2\Lambda_{\rm{v}}g^2_{\omega}g^2_{\rho}{\rho}^2 \omega
&=&g_{\omega}\left( n_{p}+n_{n}\right) ,
\label{eq:eqmw} \\
m_{\rho }^{2}{\rho}
+2\Lambda_{\rm{v}}g^2_{\omega}g^2_{\rho}{\omega}^2{\rho}
&=&\frac{g_{\rho }}{2}\left(n_{p}-n_{n}\right) ,
\label{eq:eqmr}
\end{eqnarray}%
where $n_i^s$ and $n_i$ denote the scalar densities and number densities
of species $i$, respectively. 
In the context of hadronic matter under $\beta$ equilibrium, 
the chemical potentials satisfy the relations $\mu_{p}=\mu_{n}-\mu_{e}$ and $\mu_{\mu}=\mu_{e}$.
The chemical potentials are given by
\begin{eqnarray}
\label{eq:mu}
\mu_i &=& \sqrt{{k_{F}^{i}}^{2}+{M^{\ast}}^2}+g_{\omega}\omega
+g_{\rho}\tau_{3}^{i}\rho, \hspace{0.5cm}  i=p,n,\\
\mu_{l} &=& \sqrt{{k_{F}^{l}}^{2}+m_{l}^{2}}, \hspace{3.0cm}  l=e,\mu,
\end{eqnarray}
where $M^{\ast}=M+g_{\sigma}{\sigma}$ denotes the effective nucleon mass.
The energy density of hadronic phase (HP)
is expressed as
\begin{eqnarray}
\varepsilon_{\rm{HP}} &=&\sum_{i=p,n}\frac{1}{\pi^2}
     \int_{0}^{k^{i}_{F}}{\sqrt{k^2+{M^{\ast}}^2}}\ k^2dk   \nonumber \\
&& + \frac{1}{2}m^2_{\sigma}{\sigma}^2+\frac{1}{3}{g_2}{\sigma}^3
     +\frac{1}{4}{g_3}{\sigma}^4  + \frac{1}{2}m^2_{\omega}{\omega}^2 \nonumber  \\
&&   + \frac{3}{4}{c_3}{\omega}^4
     + \frac{1}{2}m^2_{\rho}{\rho}^2
     + 3{\Lambda}_{\textrm{v}}\left(g^2_{\omega}{\omega}^2\right)
     \left(g^2_{\rho}{\rho}^2\right) \nonumber  \\
&& + \sum_{l=e,\mu}\frac{1}{\pi^{2}}\int_{0}^{k_{F}^{l}}
     \sqrt{k^{2}+m_{l}^{2}}\ k^{2}dk,
     \label{eq:ehp}
\end{eqnarray}
and the pressure is 
\begin{eqnarray}
P_{\rm{HP}} &=& \sum_{i=p,n}\frac{1}{3\pi^2}\int_{0}^{k^{i}_{F}}
      \frac{k^4dk}{\sqrt{k^2+{M^{\ast}}^2}}   \nonumber  \\
&& - \frac{1}{2}m^2_{\sigma}{\sigma}^2-\frac{1}{3}{g_2}{\sigma}^3
     -\frac{1}{4}{g_3}{\sigma}^4  + \frac{1}{2}m^2_{\omega}{\omega}^2 \nonumber \\
&&    +\frac{1}{4}{c_3}{\omega}^4
      + \frac{1}{2}m^2_{\rho}{\rho}^2 +
      \Lambda_{\textrm{v}}\left(g^2_{\omega}{\omega}^2\right)
      \left(g^2_{\rho}{\rho}^2\right) \nonumber \\
&& + \sum_{l=e,\mu}\frac{1}{3\pi^{2}}\int_{0}^{k_{F}^{l}}
     \frac{k^{4} dk}{\sqrt{k^{2}+m_{l}^{2}}}.
\label{eq:php}
\end{eqnarray}

In order to investigate the impact of EOS on hybrid star structures and the hadron-quark phase transition, we adopt several successful RMF models,
including NL3L-50~\cite{Wu2021}, BigApple~\cite{Fattoyev2020}, TM1e~\cite{Bao2014,Wu2018} and NL3~\cite{Lalazissis1997} to characterize nuclear interactions, 
which can support at least 2~$M_\odot$ NSs.
In particular, the NL3L-50 model incorporates a density-dependent coupling for the $\rho$ meson,
$g_{\rho}(n_b)=g_{\rho}(n_0)\exp\left[-a_{\rho}\left(\frac{n_b}{n_0}-1\right)\right]$,
which is varied to adjust the density dependence of the symmetry energy.
This leads to a rearrangement item $\Sigma_{r}=\frac{1}{2}\sum_{i=p,n}\frac{\partial{g_{\rho}(n_b)}}{\partial{n_b}}\tau_3{n_i}{\rho}$ for the chemical potential and pressure~\cite{Wu2021}.
The nucleon couplings in this set of EOS models are determined to reproduce the binding energies, charge radii, neutron radii of selected nuclei, and the properties of saturation nuclear matter.
These models exhibit varying stiffness in order to make our results more general, 
in which NL3L-50, BigApple and NL3 produce stiffer EOSs, while TM1e generates a relatively softer EOS.
Our choice of these parametrizations is primarily motivated by their ability to satisfy the constraint of $2~M_\odot$ and the radius constraints for NSs. 
Notably, when considering the quark degrees of freedom, a decrease in the maximum mass of the hybrid stars is expected.
Among these parameter sets, NL3L-50, BigApple and TM1e yield mass-radius relations that satisfy the constraints outlined in the introduction, therefore we consider them as representative models. 
The NL3 parameter set predicts a radius on the mass-radius curve that exceeds the established constraints, 
which is concluded to investigate the impact of this larger radius on the mixed phase core.
For completeness, we present the parameter sets of these models in Tables~\ref{tab:para}.

\begin{table}[htbp]
\caption{Masses of nucleons and mesons and meson coupling constants.
The masses are provided in units of $\mathrm{MeV}$.}
\centering
\setlength{\tabcolsep}{2.4mm}{
\begin{tabular}{lcccccccc}
\hline\hline
Parameters   &NL3L-50 &BigApple   &TM1e  &NL3   \\
\hline
$M$            &939.0    &939.0    &938.0    &939.0 \\
$m_\sigma$     &508.194  &492.730  &511.198  &508.194 \\
$m_\omega$     &782.501  &782.500  &783.000  &782.501 \\
$m_\rho$       &763.0    &763.0    &770.0    &763.0  \\
$g_\sigma$     &10.217   &9.6699   &10.0289  &10.217 \\
$g_\omega$     &12.868   &12.316   &12.6139  &12.868 \\
$g_\rho$       &8.948  &14.1618    &12.2413  &8.948  \\
$g_2/\rm{fm}^{-1}$    &10.431  &11.9214   &7.2325   &10.431  \\
$g_3$           &-28.885 &-31.6796  &0.6183   &-28.885 \\
$c_{3}$        &0       &2.6843        &71.3075  &0   \\
$\Lambda_{\rm{v}}$      &0     &0.0475        &0.0327   &0   \\
$a_\rho$         &0.583455   &0  &0        &0   \\
\hline\hline
\label{tab:para}
\end{tabular}}
\end{table}


\subsection{Quark matter phase}
\label{subsec:qm}

To describe the quark matter, we chose the SU(3) NJL model, which incorporates three flavors of quarks $u$, $d$ and $s$.
The Lagrangian density of the NJL model is given by
\begin{eqnarray}
\label{eq:Lnjl}
\mathcal{L}_{\rm{NJL}} &=&\bar{q}\left( i\gamma _{\mu }\partial ^{\mu
}-m^{0}\right) q \nonumber \\
&&+{G_S}\sum\limits_{a = 0}^8 {\left[ {{{\left( {\bar q{\lambda _a}q} \right)}^2}
+ {{\left( {\bar q i{\gamma _5}{\lambda _a}q} \right)}^2}} \right]}  \nonumber \\
&&-K\left\{ \det \left[ \bar{q}\left( 1+\gamma _{5}\right) q\right] +\det %
\left[ \bar{q}\left( 1-\gamma _{5}\right) q\right] \right\} \nonumber \\
&&- {G_V}\sum\limits_{a = 0}^8 {\left[ {{{\left( {\bar q{\gamma ^\mu }{\lambda _a}q} \right)}^2}
+ {{\left( {\bar q{\gamma ^\mu }{\gamma _5}{\lambda _a}q} \right)}^2}} \right]},
\end{eqnarray}%
where $q$ refers to the quark field, which contains three flavors ($N_f=3$) and three colors ($N_c=3$).
The current quark mass matrix is given by
$m^{0}=\text{diag} \left(m_{u}^{0},m_{d}^{0},m_{s}^{0}\right)$.
In this study, we take into account chirally symmetric four-quark interaction characterized by the coupling constant ${G_S}$,
Kobayashi--Maskawa--'t Hooft (KMT) six-quark interaction represented by the coupling constant ${K}$, 
and repulsive vector interaction governed by the coupling constant ${G_V}$. 
The inclusion of the vector coupling is essential in describing massive stars, as shown in 
Refs.~\cite{Yasutake2014,Masuda2013,Chu2015,Klahn2015,Pereiar2016,Liu2023,Alaverdyan2022}.
In this study, we adopt the parameters provided in Ref.~\cite{Rehberg1996,Ruester2005},
$m_{u}^{0}=m_{d}^{0}=5.5\ \text{MeV}$, $m_{s}^{0}=140.7\ \text{MeV}$,
$\Lambda =602.3\ \text{MeV}$, ${G_S}\Lambda^{2}=1.835$,
and $K\Lambda ^{5}=12.36$. These parameters are fitted to the pseudoscalar meson octet in vacuum.
In a recent study, Gholami et al.~\cite{Gholami2024b} generalized an idea based on the requirement of renormalization-group (RG) consistency to expand the scope of the cutoff $\Lambda$.
Through this approach, we can examine the effect of the scalar coupling. 
In this study, we introduce $\lambda=\frac{\Lambda^{'}}{\Lambda}$, where $\Lambda^{'}$ is larger cutoff with RG consistency . 
The new scalar coupling then becomes $G_S^{'}=\frac{1}{\lambda^2}G_S$.
The vector coupling ${G_V}$ is treated as a free parameter in our analysis, 
following the approach adopted in Refs.~\cite{Yasutake2014,Pereiar2016,Wu2018,Wu2019},
because there is no well-constrained value for only ${G_V}$ at finite density currently.
By treating ${G_V}$ as a free parameter, we aim to explore its influence on the properties and behavior of quark matter within our study.
Since the vector coupling  ${G_V}$ only serves to stiffen the EOS of quark matter, 
its effects on the hadron-quark phase transition are expected to be qualitatively similar across different models~\cite{Klahn2015,Wu2019,Alaverdyan2022}. 
By including  ${G_V}$,  the EOS would become more resistant to compression, resulting in increased pressure for a given density.

At the mean-field level, the constituent quark masses arise from spontaneous chiral symmetry breaking.
In vacuum, the constituent quark mass $m_{i}^{\ast}$ is much larger than the current quark mass $m_{i}^{0}$.
The determination of constituent quark masses  $m_{i}^{\ast}$ in quark matter involves solving the relevant gap equations,
\begin{equation}
\label{eq:gap}
m_{i}^{\ast }=m_{i}^{0}-4{G_S}\langle \bar{q}_{i}q_{i}\rangle +2K\langle \bar{q}%
_{j}q_{j}\rangle \langle \bar{q}_{k}q_{k}\rangle,
\end{equation}%
with ($i,j,k$) being any permutation of ($u,d,s$).
The energy density of quark matter is given by
\begin{eqnarray}
\label{eq:enjl}
\varepsilon_{\rm{NJL}} &=&\sum\limits_{i = u,d,s}
 {\left[ { - \frac{3}{{{\pi ^2}}}\int_{k_F^i}^\Lambda
  {\sqrt {{k^2} + m_i^{ * 2}} } \;{k^2}dk} \right]} \notag\\
 & &  + 2{G_S}\left( {C_u^2 + C_d^2 + C_s^2} \right) - 4K{C_u}{C_d}{C_s} \notag\\
 & &  + 2{G_V}\left( {n_u^2 + n_d^2 + n_s^2} \right)   - {\varepsilon _0},
\end{eqnarray}%
where $C_{i}=\left\langle \bar{q}_{i}q_{i}\right\rangle $ denotes the quark condensate
of flavor $i$.  
The constant $\varepsilon_{0}$ is introduced to ensure that the energy density in the physical vacuum is zero. 
In our current study, the choice of
$\varepsilon_{0}$ leads the pressure also vanishes in the vacuum. 
In quark matter, the chemical potentials of quarks and leptons satisfy the $\beta$ equilibrium condition, which is expressed as
$\mu_{s}=\mu_{d}=\mu_{u}+\mu_{e}$ and $\mu_{\mu}=\mu_{e}$. 
The chemical potential of quark flavor $i=u,d,s$ is given by
\begin{eqnarray}
\mu_i =\sqrt{{k_{F}^{i}}^{2}+{m_{i}^{\ast}}^{2}}+ 4 G_V  n_i.
\end{eqnarray}
The total energy density and pressure in quark phase (QP)
are written as
\begin{eqnarray}
\label{eq:eqp}
\varepsilon_{\rm{QP}} &=& \varepsilon_{\rm{NJL}}
  +\sum_{l=e,\mu }\frac{1}{\pi^{2}}
\int_{0}^{k_{F}^{l}}\sqrt{k^{2}+m_{l}^{2}}\ k^{2}dk,
\notag\\
P_{\rm{QP}} &=&\sum_{i=u,d,s,e,\mu }n_{i}\mu_{i}-\varepsilon_{\rm{QP}}.
\label{eq:pqp}
\end{eqnarray}

\subsection{Hadron-quark mixed phase}
In our study, we employ the Gibbs construction to describe the hadron-quark mixed phase. 
Within this construction, the system satisfies the $\beta$ equilibrium. 
Both hadronic matter and quark matter
are allowed to be charged separately, but the total charge remains zero.
The energy density of the
mixed phase (MP) is 
\begin{equation}
\varepsilon_{\rm{MP}} = u \varepsilon_{\rm{QP}}
  + (1 - u)\varepsilon_{\rm{HP}},
\label{eq:emix}
\end{equation}
where $u$ is the volume fraction of quark matter.
The pressure equilibrium and the chemical potential equilibrium between two phases are shown below,
\begin{eqnarray}
P_{\rm{HP}} &=& P_{\rm{QP}}, \label{eq:CP2} \\
\mu_u+\mu_e &=& \mu_d = \mu_s = \frac{1}{3}\mu_n + \frac{1}{3}\mu_e. 
\end{eqnarray}
At a given baryon density $n_b$, there are two independent chemical potentials, $\mu_{n}$ and $\mu_{e}$, 
which can be determined by the constraints of global charge neutrality and baryon number conservation given in 
\begin{eqnarray}
\label{eq:nc}
0 &=& n_e + n_\mu -\frac{u}{3}\left( 2n_u - n_d - n_s \right)
  - (1 - u)  n_p   , \\
n_b &=& \frac{u}{3}\left( n_u + n_d + n_s \right)
    + (1 - u) \left( n_p + n_n \right).
\label{eq:nb}
\end{eqnarray}
All the properties of the mixed phase can be calculated 
under the equilibrium state with given $n_b$.

\section{Results and Discussion}
\label{sec:results}

\begin{figure*}[!htb]
\includegraphics[viewport=20 10 580 580, scale=0.3]{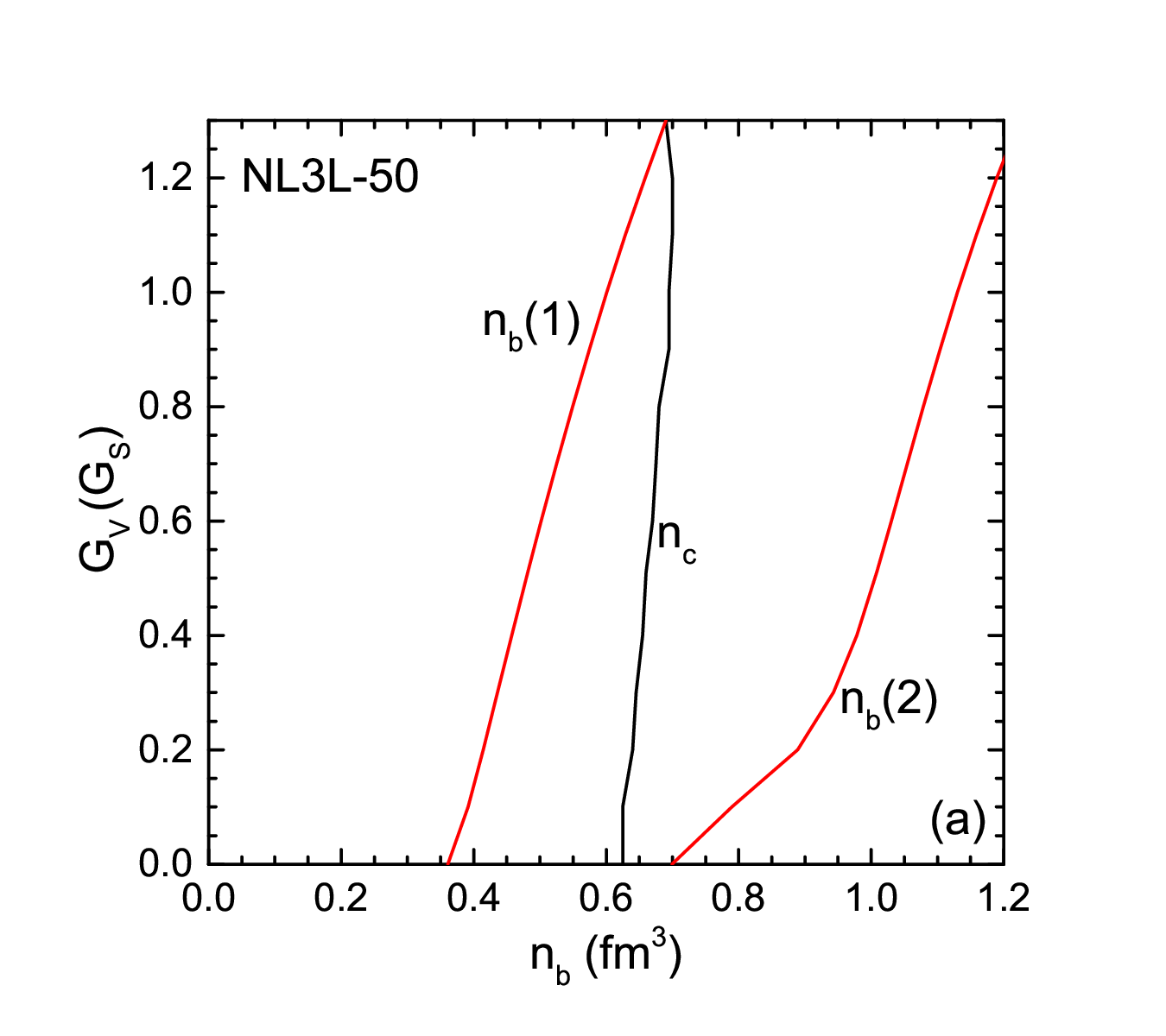}%
\includegraphics[viewport=20 10 580 580, scale=0.3]{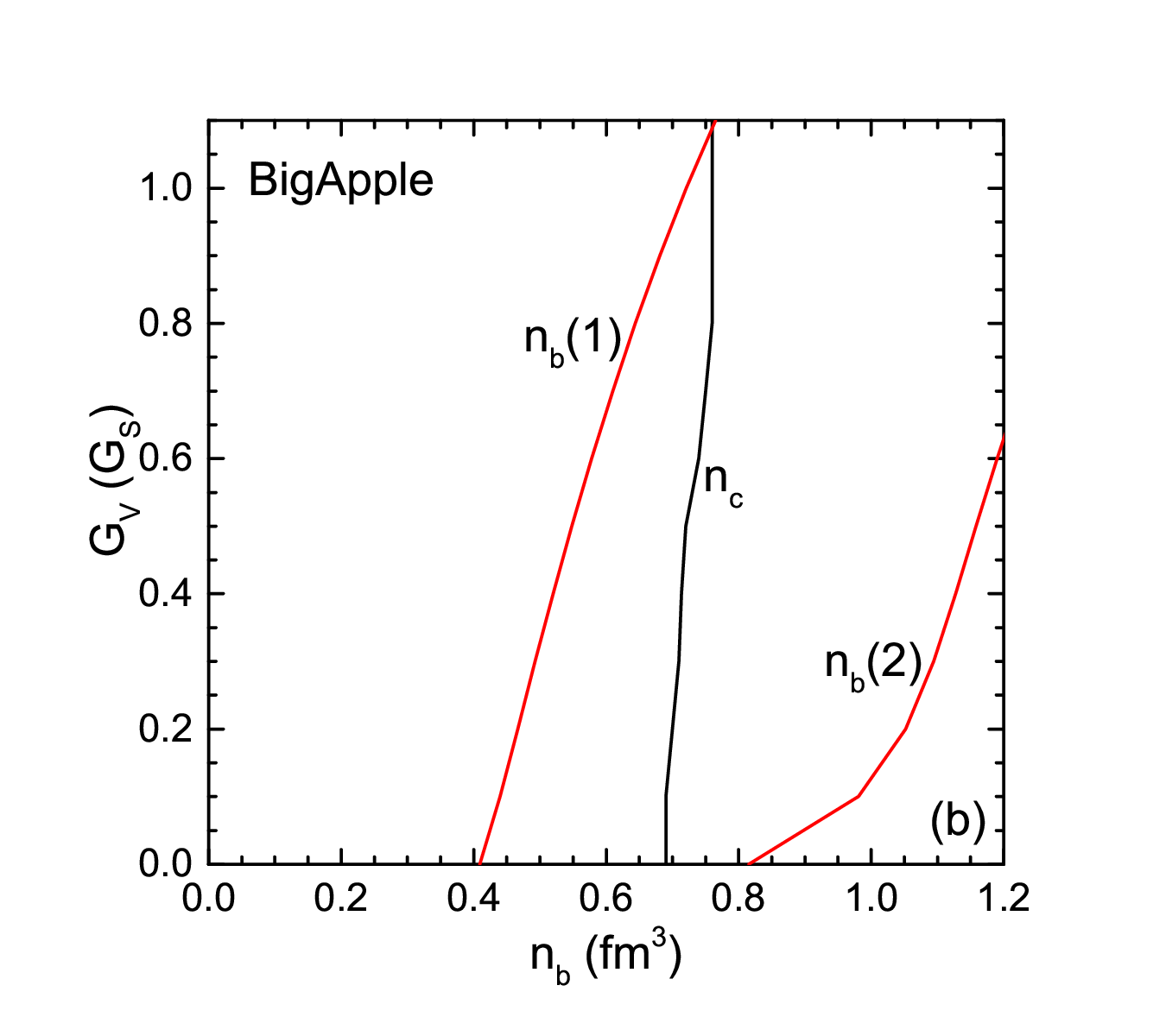}%
\includegraphics[viewport=20 10 580 580, scale=0.3]{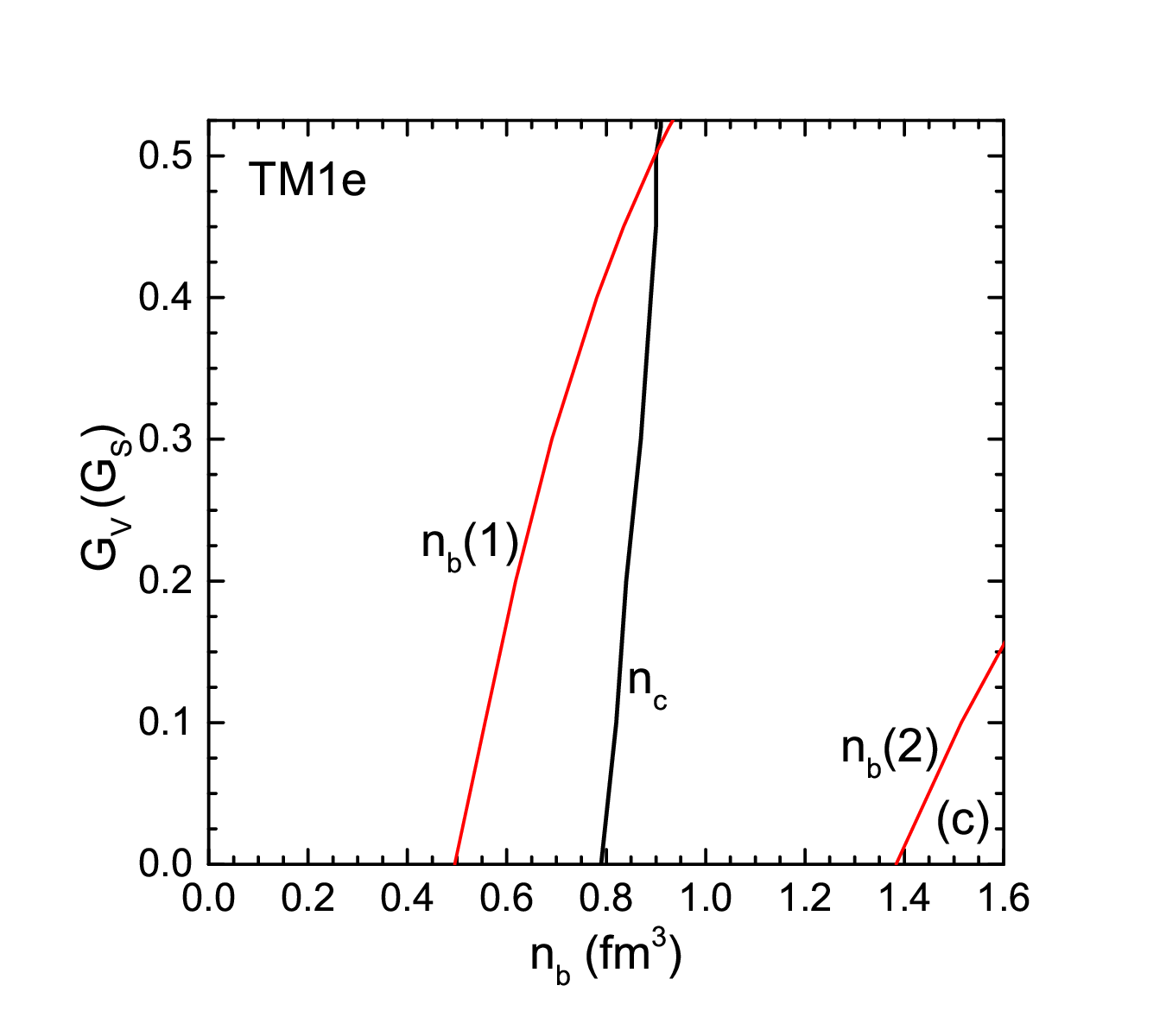}%
\caption{The phase transition density $n_b(1)$, $n_b(2)$ and the center density $n_c$ of the maximum-mass NS as a function of baryon number density $n_b$ with different values of $G_V$.}
\label{fig:1gvnb}
\end{figure*}

The presence of deconfined quarks in the core of massive hybrid stars is an intriguing possibility. 
In this section, we investigate the EOS and its impact on the internal structure of the maximum-mass hybrid stars. 
To describe the hadronic matter, we employ the RMF models, 
while the NJL model with repulsive vector coupling is utilized for quark matter. 
The hadron-quark mixed phase is treated under the Gibbs equilibrium condition. 
For the representation of larger mass NSs, we employ the stiff parameter sets NL3L-50 and BigApple,
while the TM1e parameter set illustrates results for NSs with a mass around $2~M_\odot$. 
The results obtained from the NL3 parameter set are included in select figures for comparison, 
as this set tends to predict excessively large radii and tidal deformability compared to observational data.

We examined the effect of vector coupling on the phase transition density and the central density $n_c$ of maximum-mass hybrid star, as shown in Fig.~\ref{fig:1gvnb}. 
With $G_V$ increase, the mixed phase is delayed to higher densities, affecting both $n_b(1)$ (the transition density between hadronic phase and the mixed phase) 
and $n_b(2)$ (the transition density between the mixed phase and quark phase),
while the central density $n_c$ remains relatively stable. 
If $n_c<n_b(1)$, it indicates that the stable central density of the maximum-mass star occurs in pure hadronic matter without deconfined quarks.
The maximum mass of hybrid star is primarily determined by the stiffness EOS of the hadronic phase, 
while the range of the hadronic phase is influenced by the vector coupling $G_V$.
Furthermore, a strong enough $G_V$ can prevent the hadron-quark phase transition, 
with the threshold value of $G_V$ being determined by the EOS of hadronic phase.
NL3L-50, BigApple, and TM1e support maximum $G_V$ values of approximately $G_V<1.3~G_S$, $G_V<1.1~G_S$, and $G_V<0.5~G_S$, respectively.
The relatively large upper limit derived here benefits the stiffness of the total EOS. 
The decrease in the range between  $n_b(1)$ and $n_c$ as $G_V$ increases indicates a reduction in the mixed phase range. 
Beyond the threshold value of $G_V$, although the hadron-quark phase transition may occur, 
the mixed phase is not supported in a hybrid star.
For a given $G_V$, the value of $n_b(1)$ derived from BigApple shifts slightly to a higher density compared to the results from NL3L-50, 
and $n_b(2)$ shifts more significantly.
This suggests that nucleons with NL3L-50 dissociate faster to quarks than with BigApple (and TM1e) interaction.
In this work we aim to explore the possibility of a sizable mixed-phase core, so we utilize values of $G_V=0,\ 0.1\ G_S,\ 0.2\ G_S$ in the following.

\begin{figure*}[htb]
\includegraphics[bb=10 10 580 400, width=0.325\linewidth]{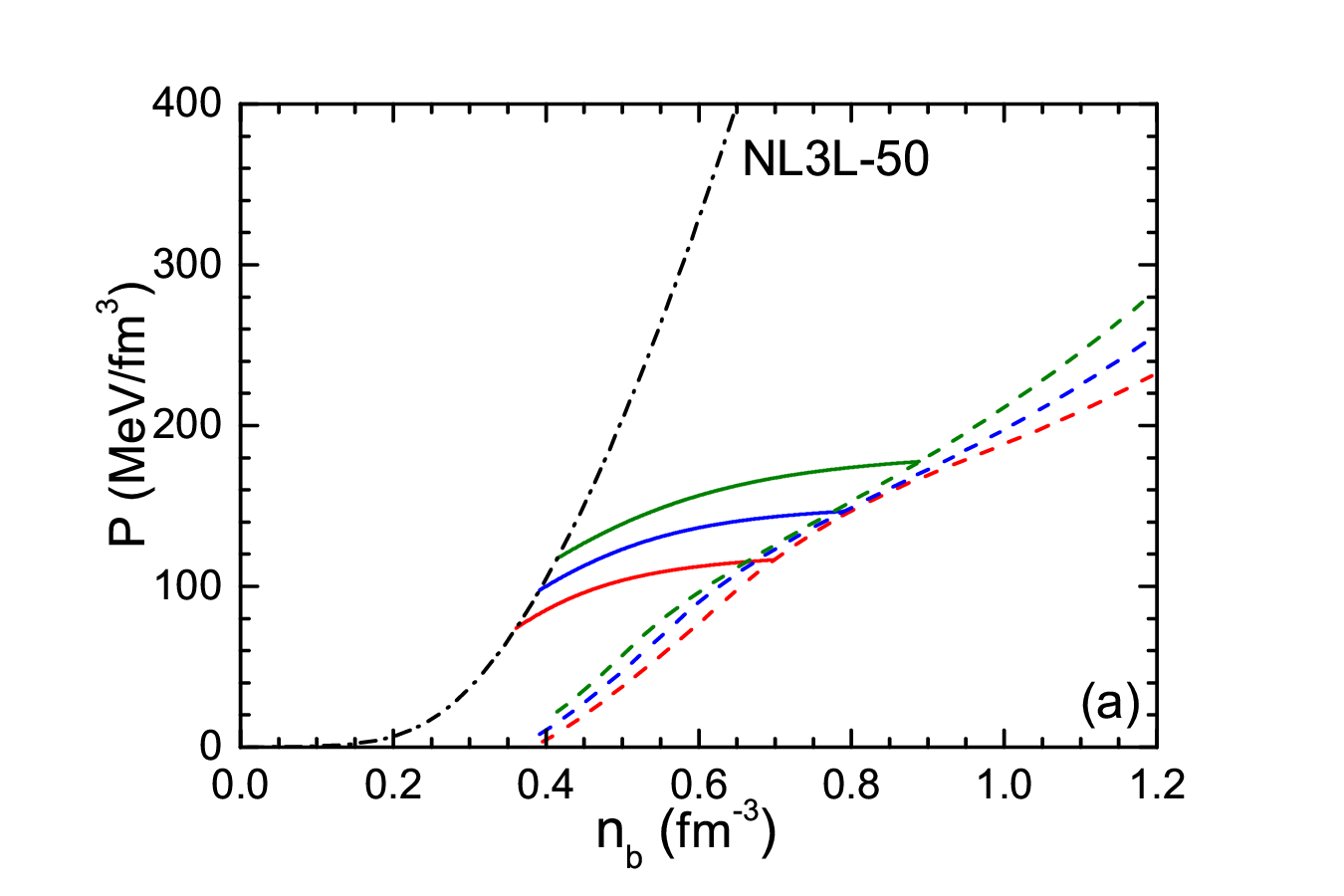}%
\includegraphics[bb=10 10 580 400, width=0.325\linewidth]{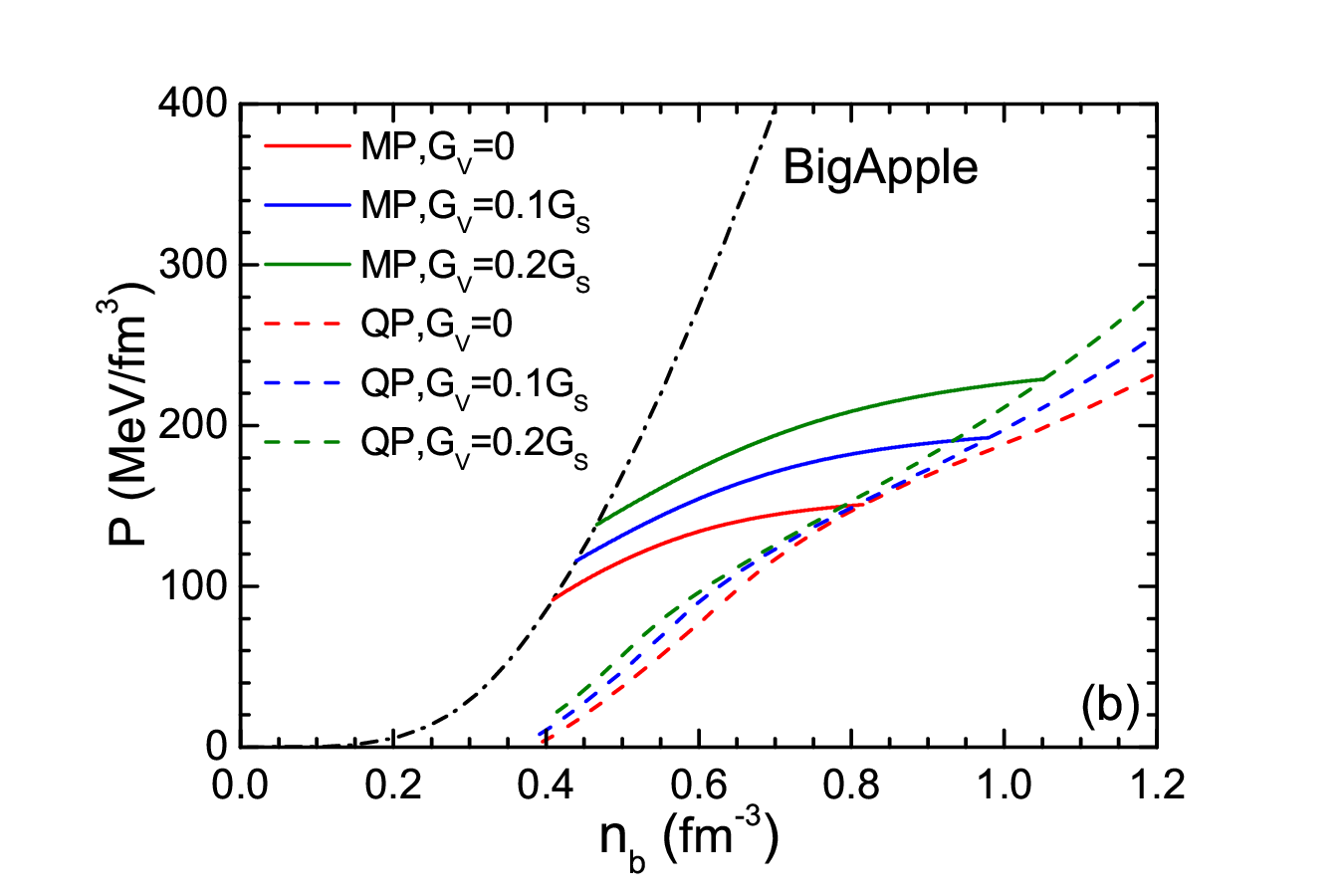}%
\includegraphics[bb=10 10 580 400, width=0.325\linewidth]{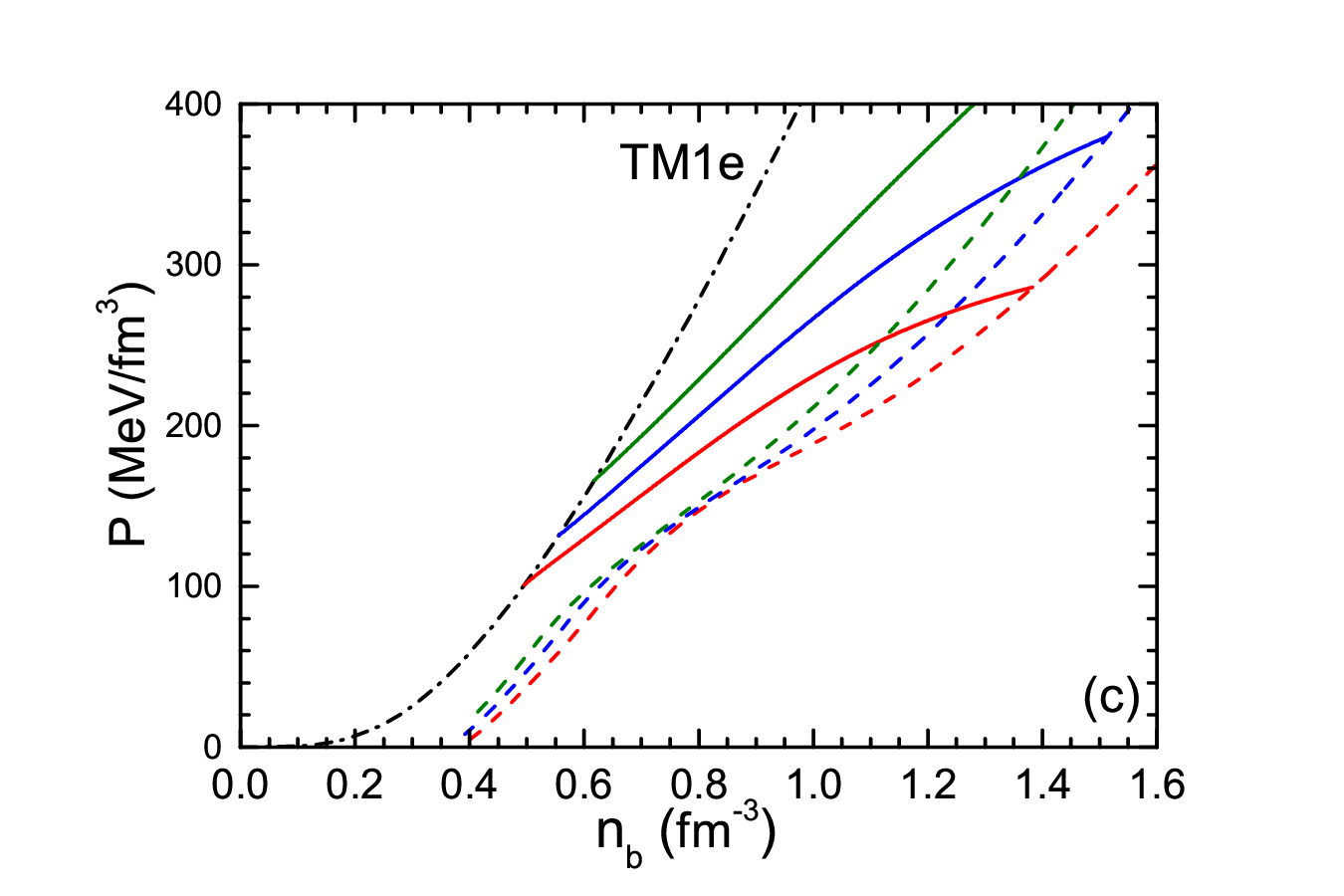}%
\caption{Pressures $P$ as a function of the number density $n_b$  obtained using different parameter sets as NL3L-50, BigApple, and TM1e. 
The results for the hadronic phase, mixed phase (MP), and quark phase (QP) are represented by dash-dot lines, solid lines and dashed lines, respectively.
Additionally, the strength of the vector coupling is varied with values $G_V=0,\ 0.1\ G_S,\ 0.2\ G_S$, which are depicted by red, blue, and green lines, respectively.}
\label{fig:2nbp}
\end{figure*}
\begin{figure*}[htb]
\includegraphics[bb=10 30 410 550, width=0.3\linewidth]{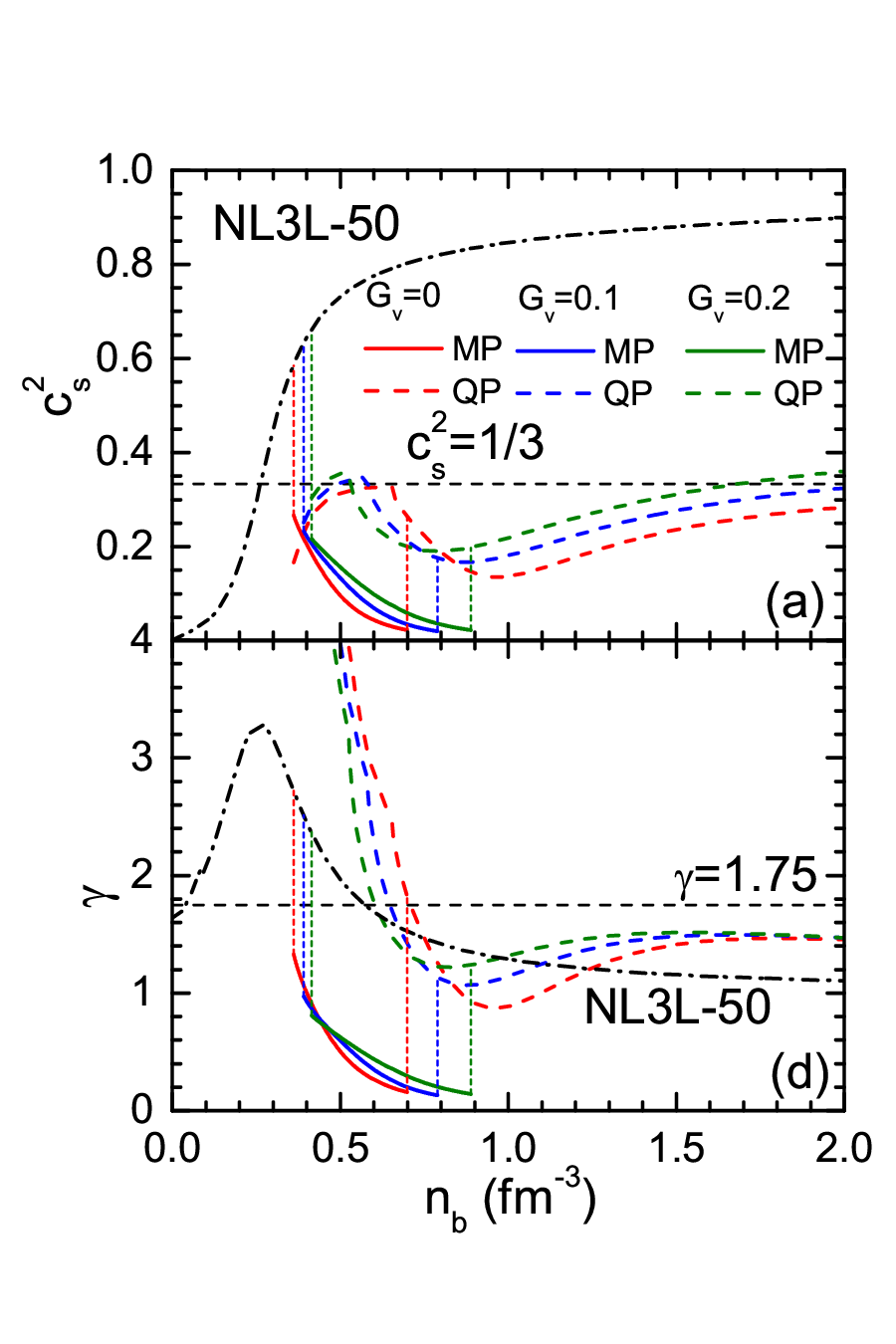}
\includegraphics[bb=10 30 410 550, width=0.3\linewidth]{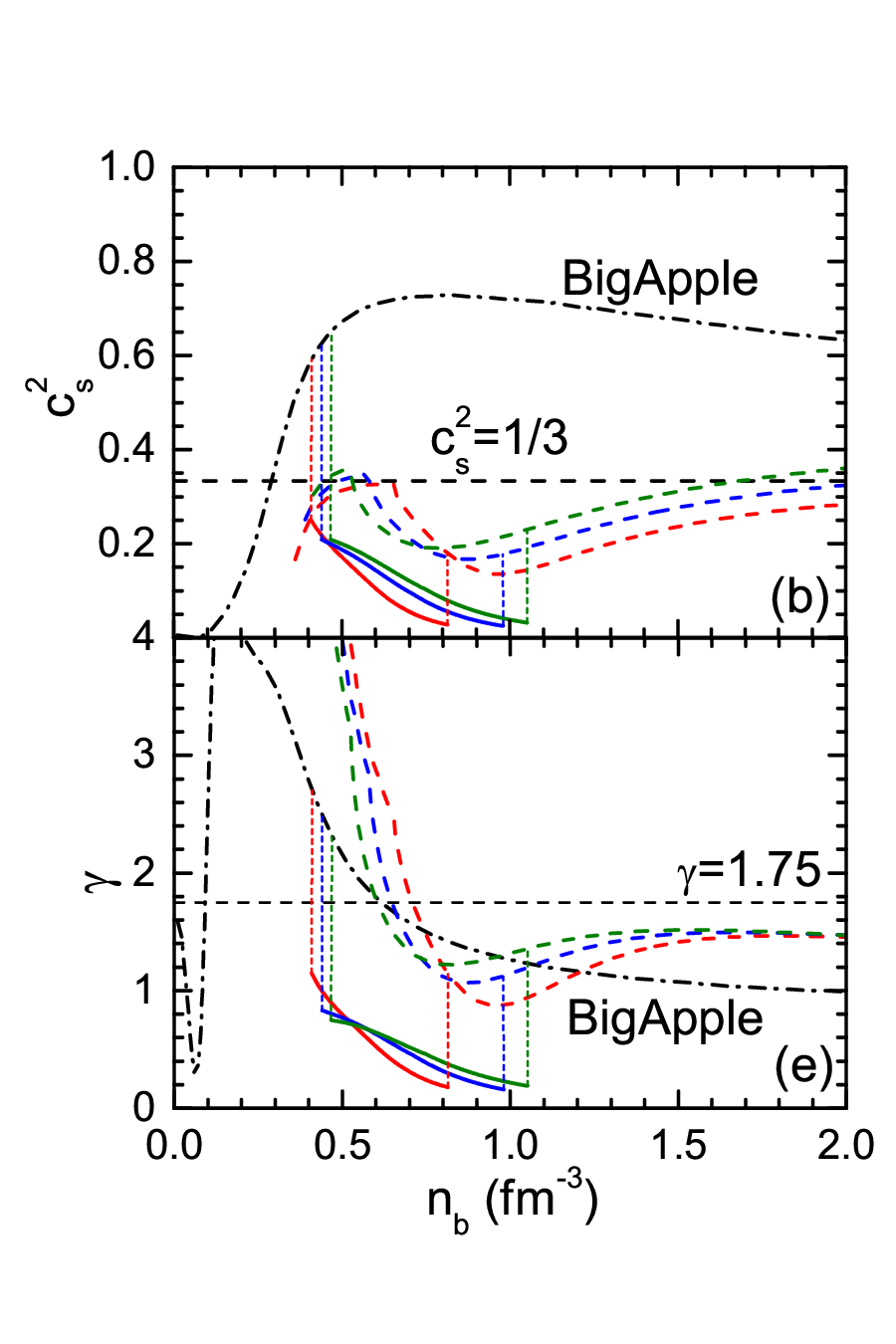}
\includegraphics[bb=10 30 410 550, width=0.3\linewidth]{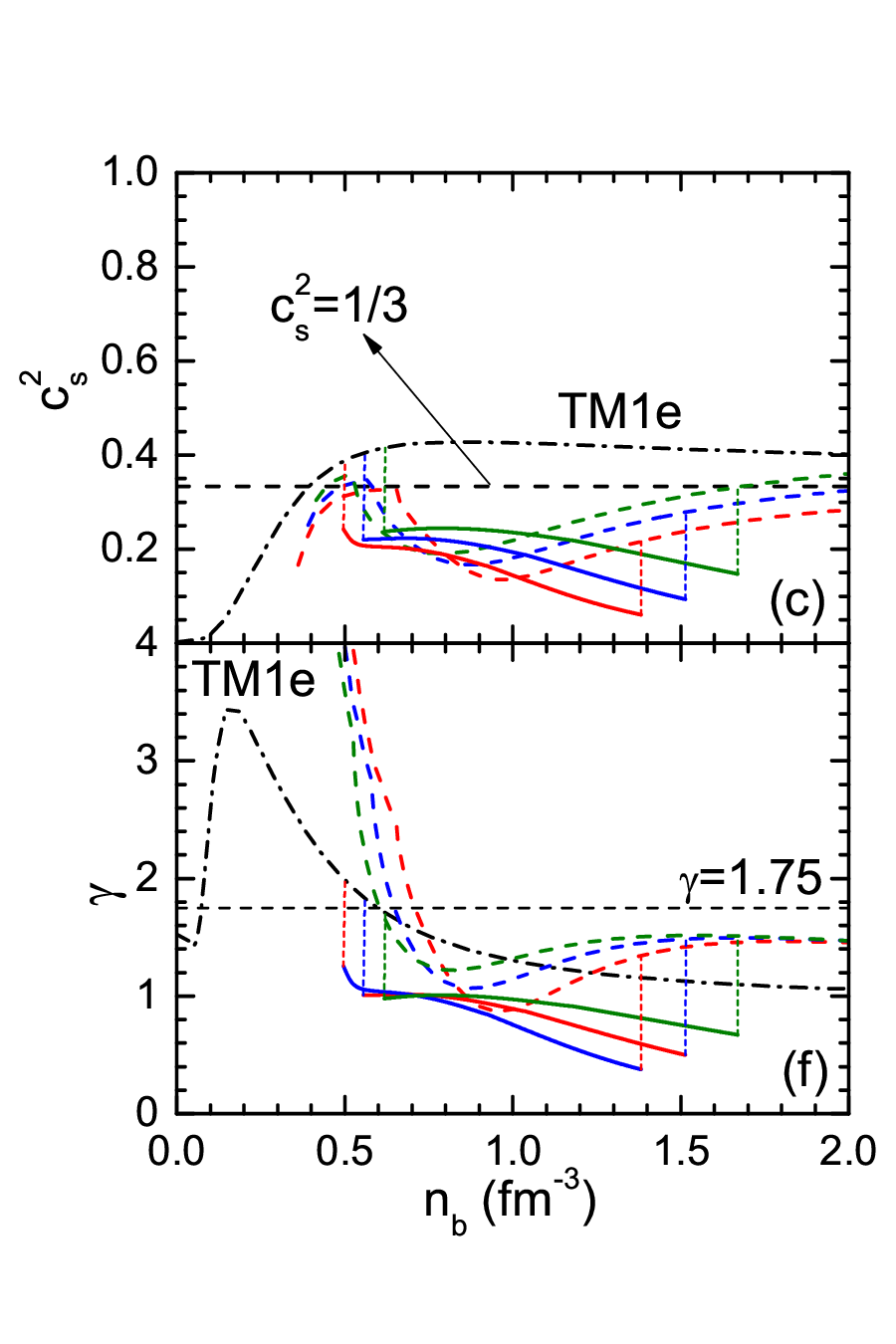}
\caption{The squared sound velocity $c_s^2$ (upper panel)
and the polytropic index $\gamma$ (lower panel) as functions of the baryon number density $n_b$ with varying vector couplings $G_V=0,\ 0.1\ G_S,\ 0.2\ G_S$. 
The labels are consistent with those in Fig.~\ref{fig:2nbp}.
Additionally, in the upper panel, a short dashed line is utilized to represent the conformal limit with $c_s^2=1/3$.
In the lower panel, $\gamma=1.75$ (short dashed line) serves as reference value to distinguish the nucleon degree of freedom from non-nuclear degrees of freedom.}
\label{fig:3nbcs2}
\end{figure*}

\begin{figure*}[!htbp]
\includegraphics[bb=10 30 540 480, scale=0.38]{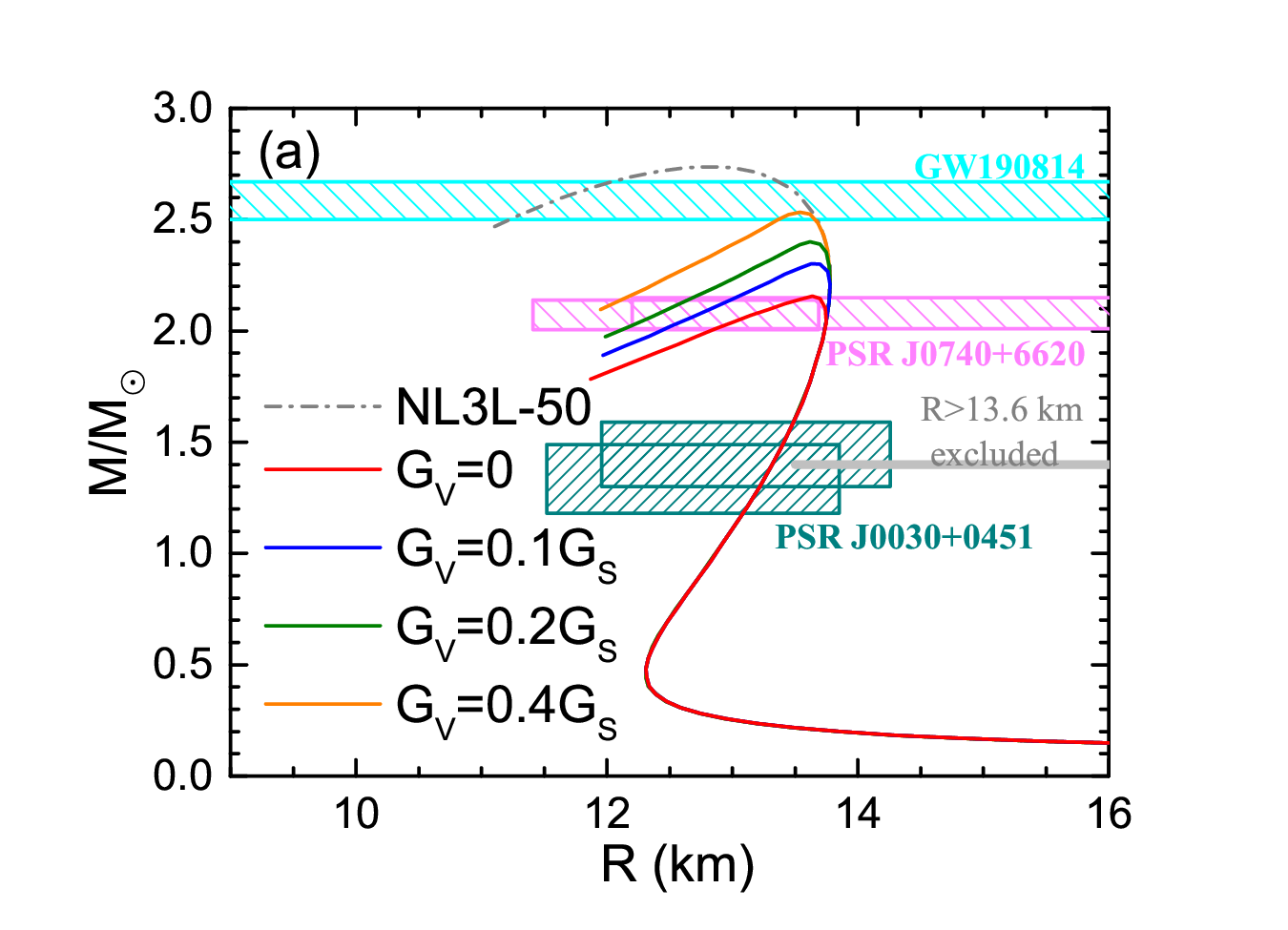}%
\includegraphics[bb=10 30 540 480, scale=0.38]{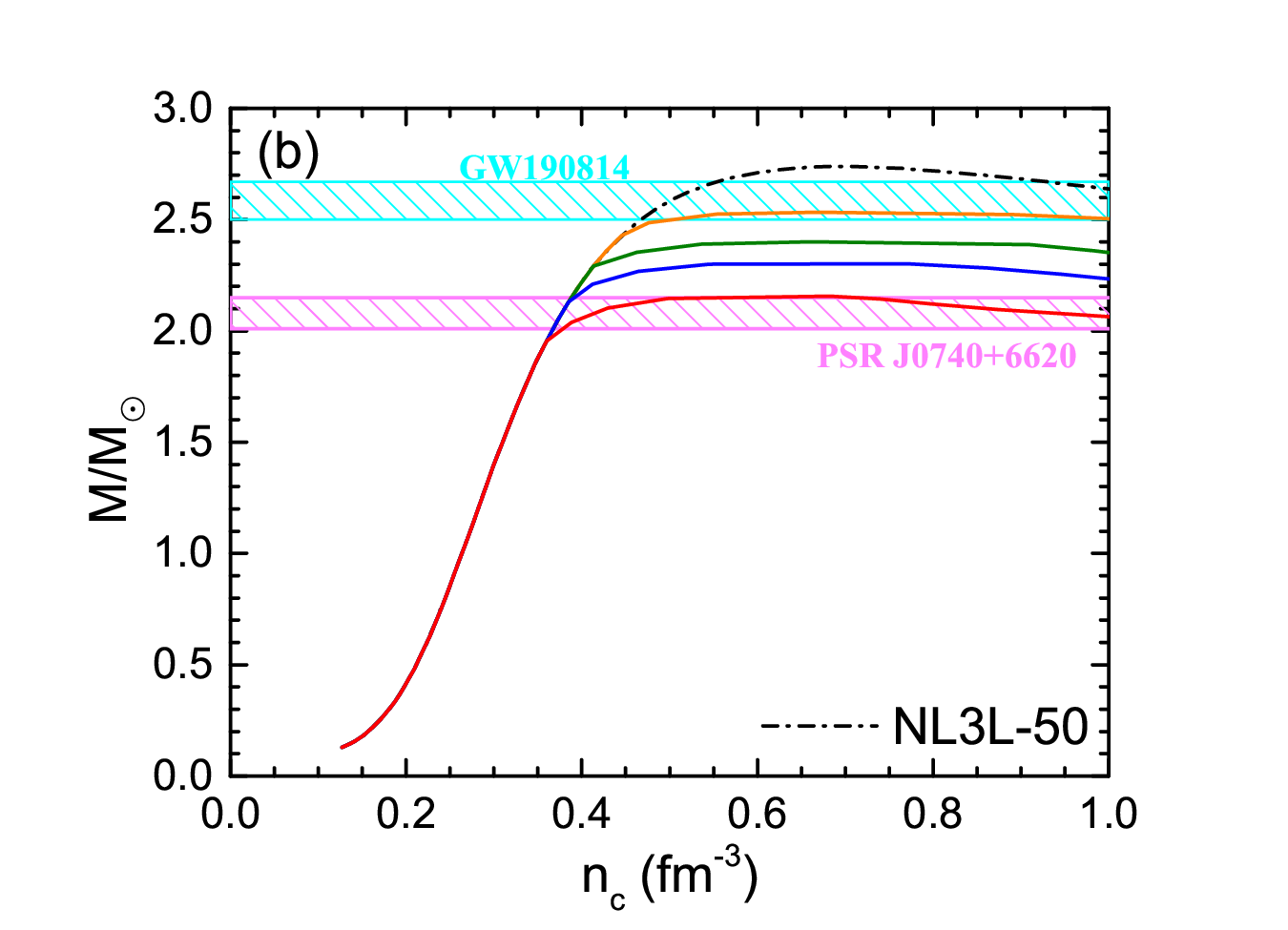}\\
\includegraphics[bb=10 30 540 480, scale=0.38]{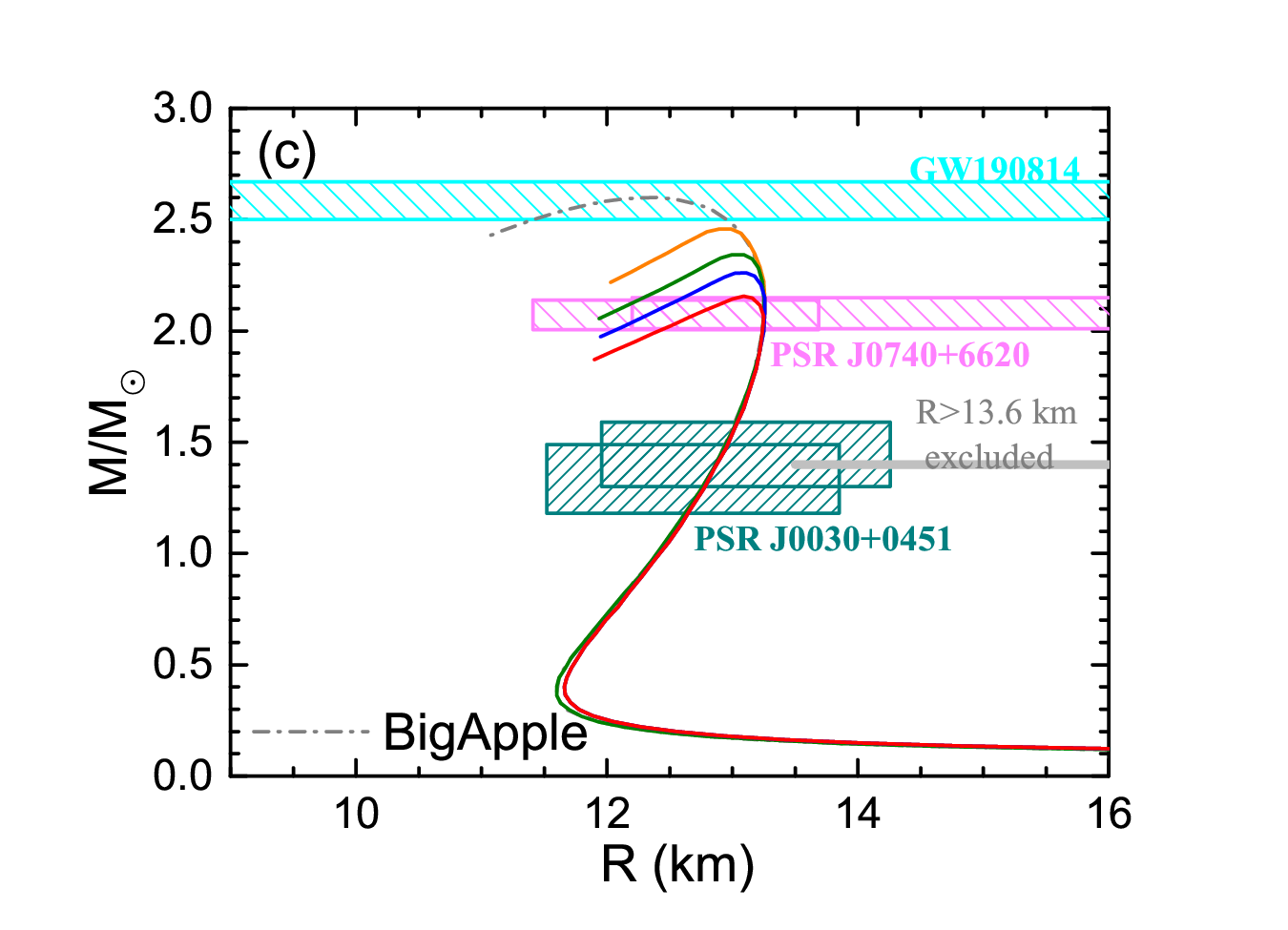}%
\includegraphics[bb=10 30 540 480, scale=0.38]{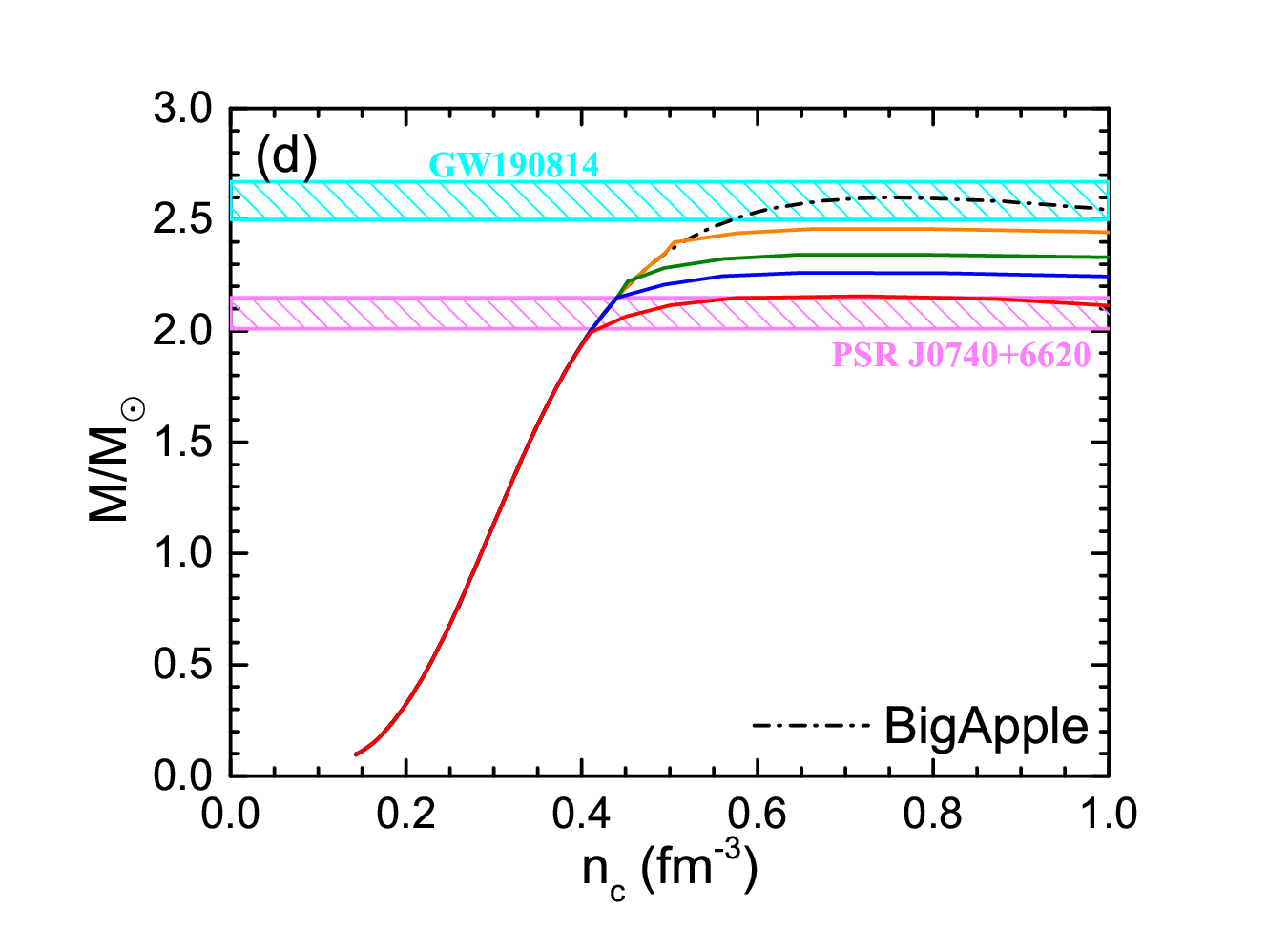}\\
\includegraphics[bb=10 30 540 480, scale=0.38]{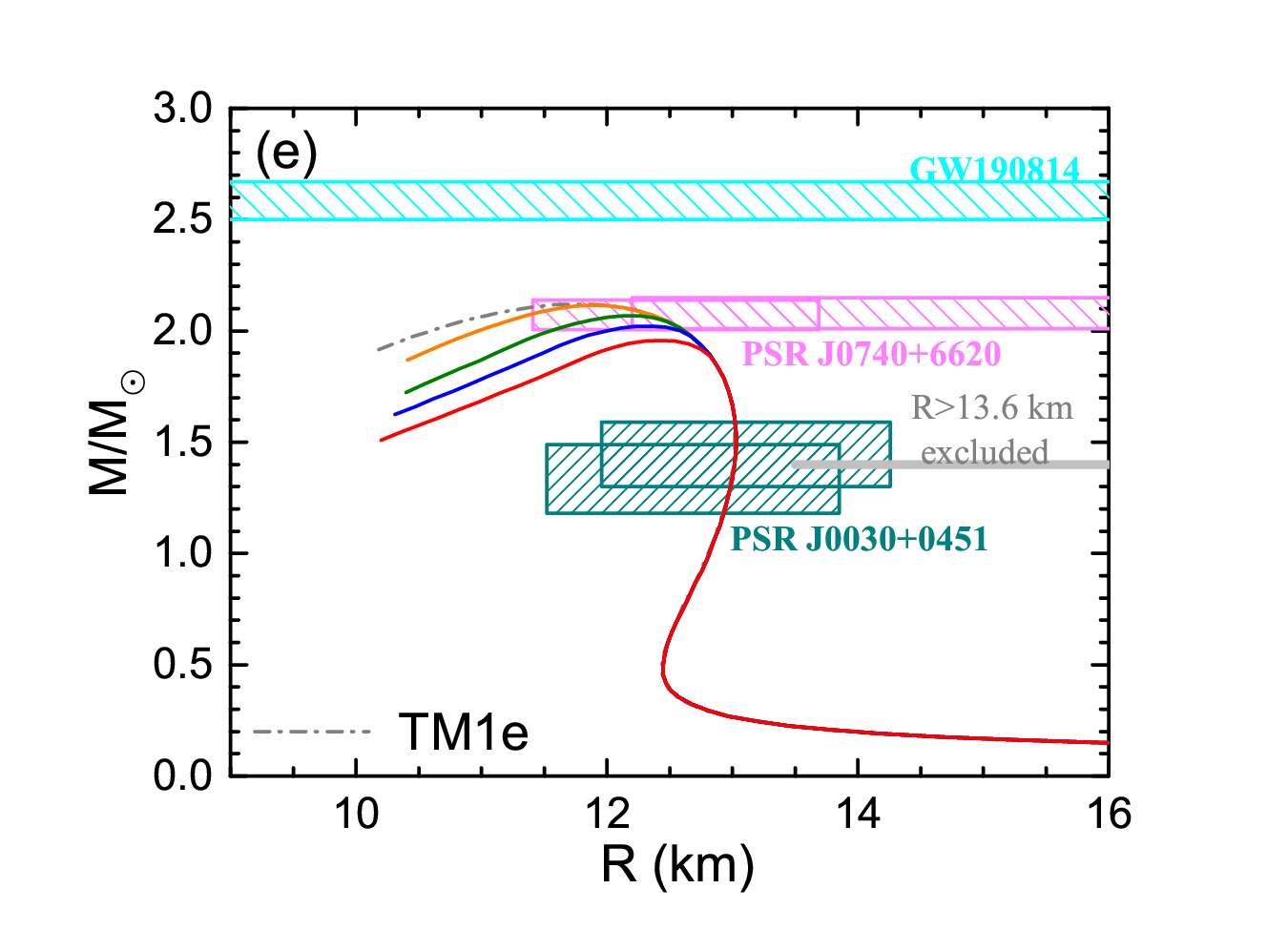}%
\includegraphics[bb=10 30 540 480, scale=0.38]{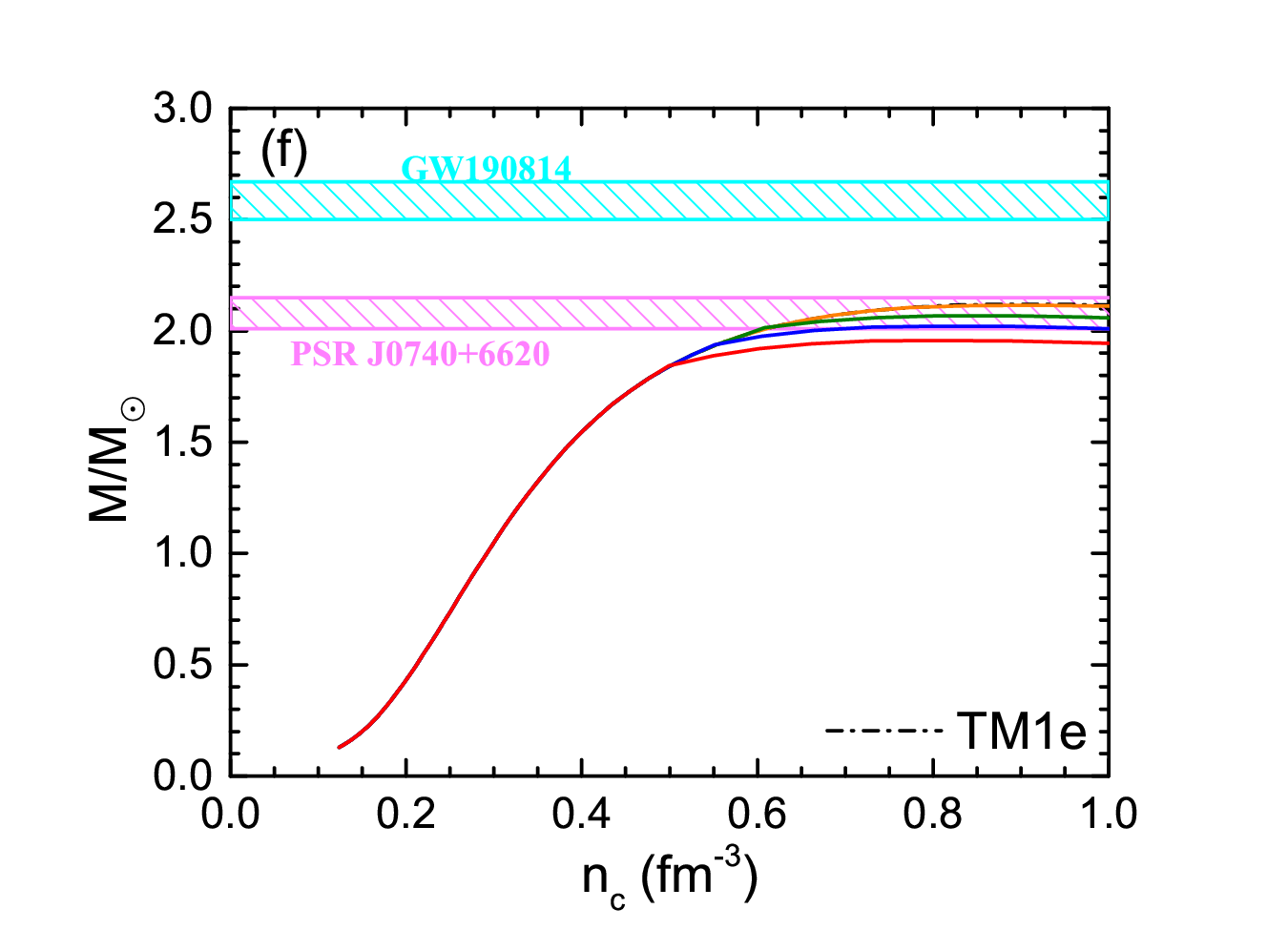}\\
\caption{The  mass-radius relations (left panel) and mass-central density $n_c$ relations (right panel) of hybrid stars with different model parameters. 
The results from pure hadronic EOS (dash-dot lines) are compared with those including hadron-quark phase transition for different vector couplings. 
The shaded areas correspond to simultaneous measurements of the mass and radius range from NICER
for PSR J0030+0451~\cite{Riley2019,Miller2019} and PSR J0740+6620~\cite{Riley2021,Miller2021}, respectively. 
The radius constraint $R_{1.4}\leq13.6~\rm{km}$ is presented with light grey~\cite{Annala2018}.
The hypothesis that the second component of GW190814 is a NS is also depicted~\cite{Abbott2020}.}
\label{fig:4rm}
\end{figure*}

\begin{figure*}[!htbp]
\includegraphics[bb=10 30 540 480, scale=0.31]{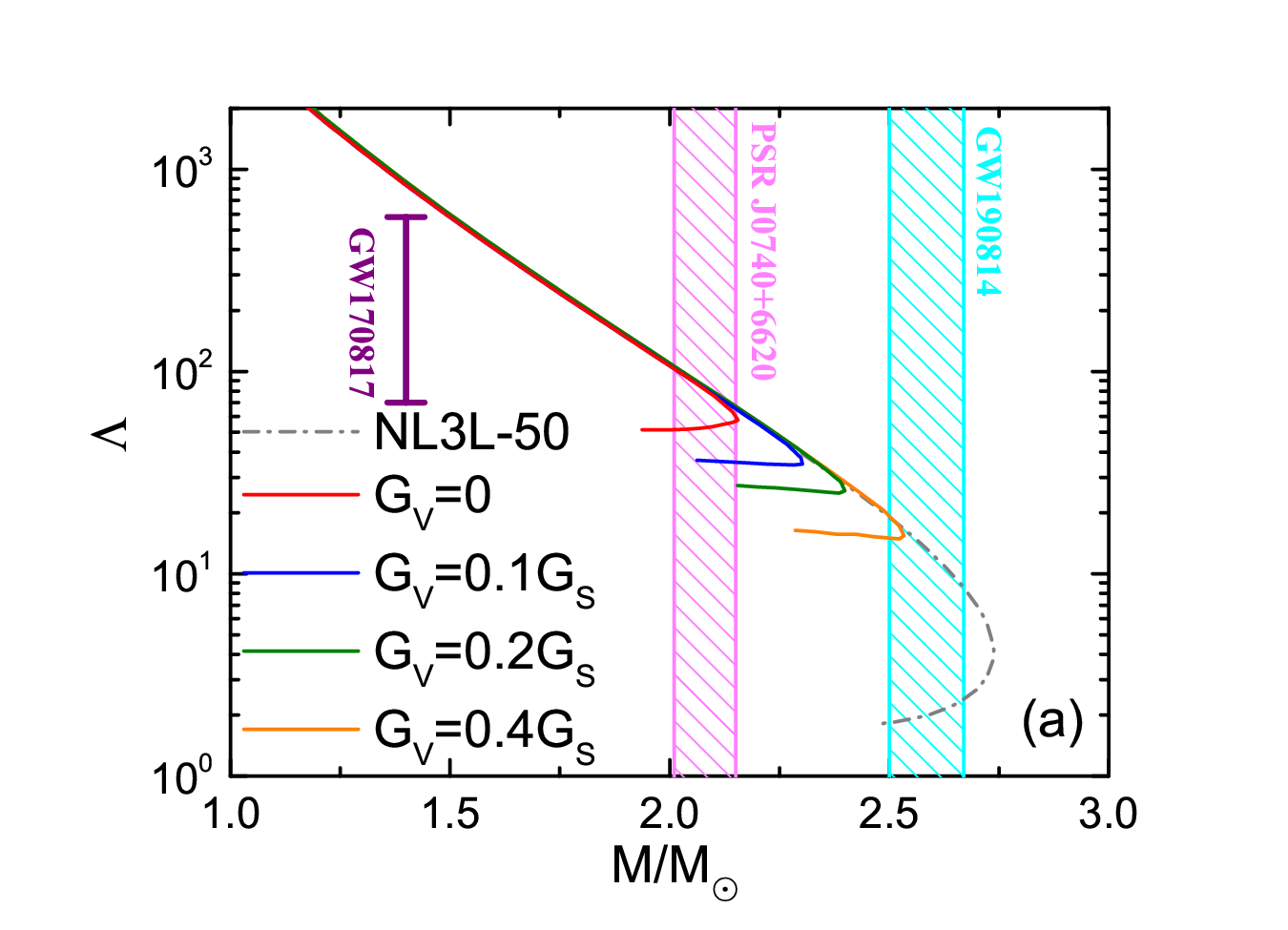}%
\includegraphics[bb=10 30 540 480, scale=0.31]{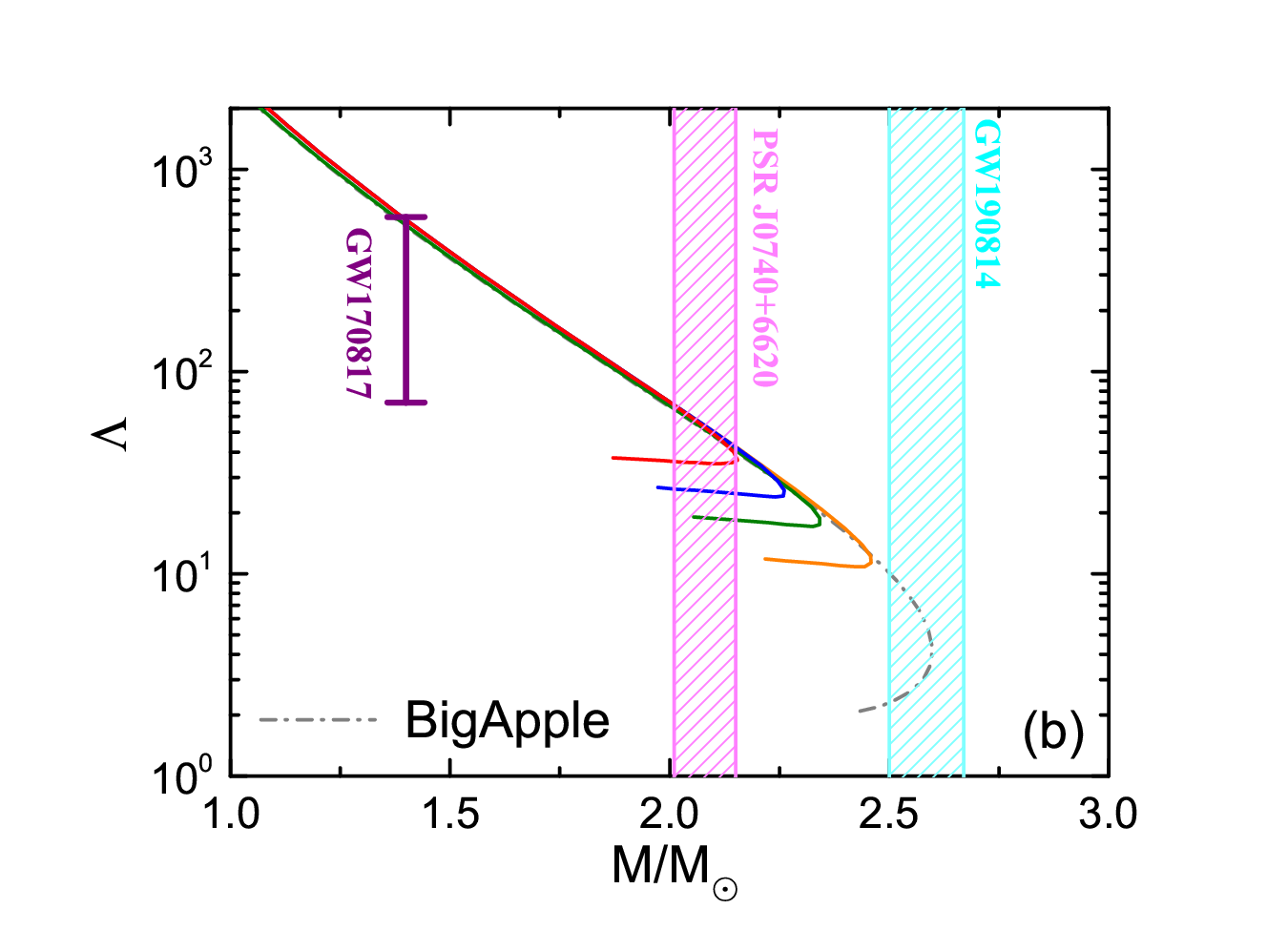}%
\includegraphics[bb=10 30 540 480, scale=0.31]{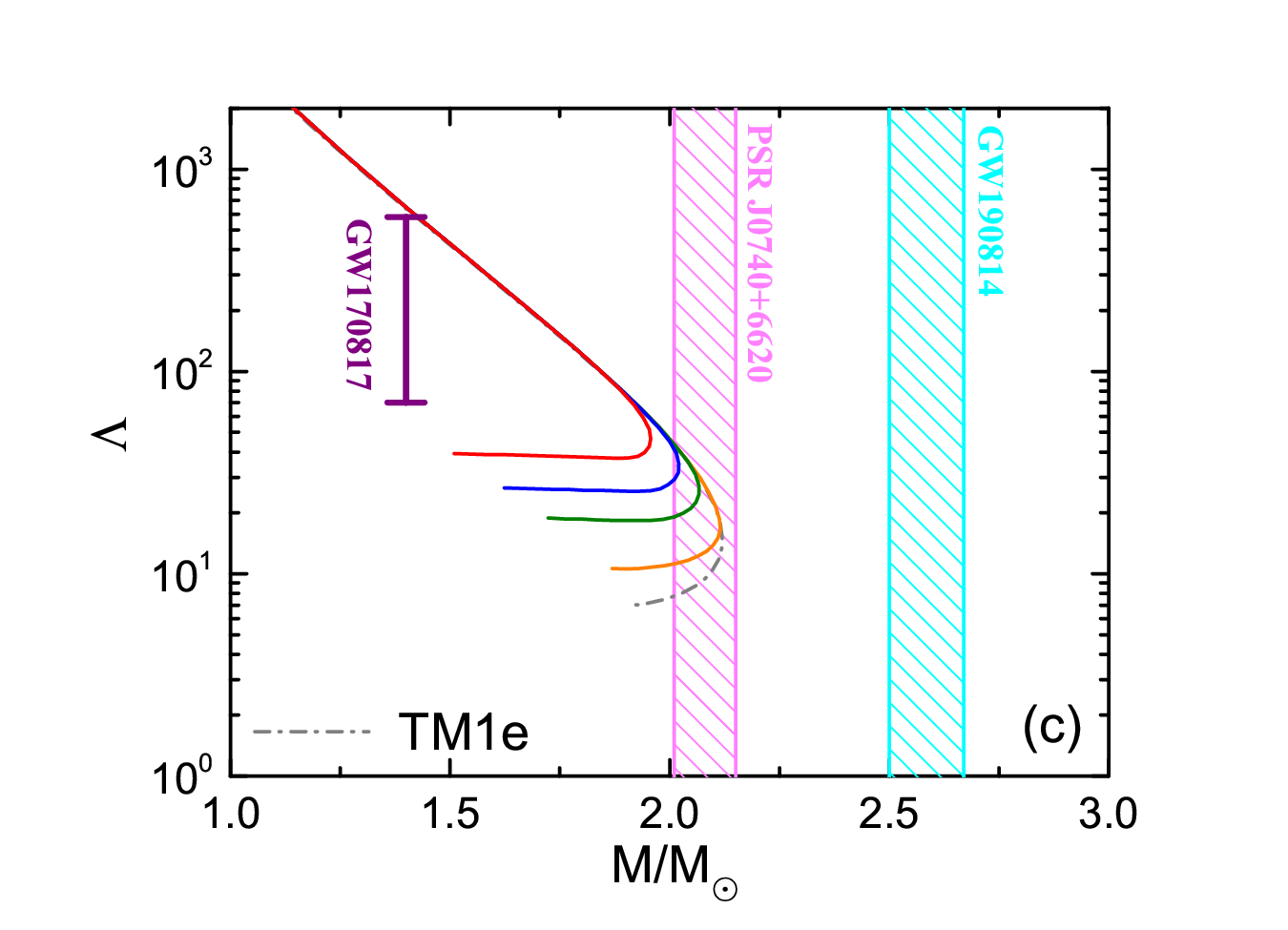}%
\caption{The  dimensionless tidal deformability as a function of NS mass.
The purple vertical line indicates the tidal deformability constraint at 1.4 $M_\odot$ from GW170817 event with $\Lambda_{1.4}=190^{+390}_{-120}$~\cite{Abbott2018}.
The shaded areas correspond to the mass constraints from PSR J0740+6620 and GW190814 event~\cite{Fonseca2021,Abbott2020}, respectively. }
\label{fig:5ml}
\end{figure*}

\begin{table*}[htbp]
\caption{Maximum-mass hybrid stars bulk properties with $G_V=0$,
 including the maximum mass $M_{\mathrm{max}}$, the corresponding radius $R_{\mathrm{max}}$, the onset density of the mixed phase $n_b$(1), the mixed phase radius $R_{\rm MP}^{\rm max}$, its fraction and the central number density $n_c$.
 The properties of 2~$M_{\odot}$ hybrid stars using $G_V=0$ is also shown, 
 including the radius $R_{2.0}$, the mixed phase radius $R_{\rm MP}^{2.0}$ and the central density $n_{c}(2.0)$.
 The last column is the radius of 1.4~$M_\odot$ NS $R_{1.4}$.}
	\begin{center}
		\setlength{\tabcolsep}{2.1mm}{
	\begin{tabular}{lcccccccccccc}
		\hline\hline
		  Model   & $M_{\rm max}$ & $R_{\rm max}$ & $n_{b}(1)$ & $R_{\rm MP}^{\rm max}$ & $R_{\rm MP}^{\rm max}/R_{\rm max}$ & $n_{c}$(max)   & $R_{2.0}$ & $R_{\rm MP}^{2.0}$ & $R_{\rm MP}^{2.0}$/$R_{2.0}$ & $n_{c}(2.0)$ & $R_{1.4}$\\
			  & $(M_{\odot})$ & (km) & ${\rm (fm^{-3})}$ & (MeV) & ($\%$) & ${\rm (fm^{-3})}$  & (km)  & (km) & ($\%$) & ${\rm (fm^{-3})}$ & (km)\\
   \hline
		NL3L-50    & 2.156 & 13.65 & 0.361 & 5.46 & 40.00 & 0.625    & 13.74 & 2.26 & 16.45 & 0.373 & 13.32\\
		BiaApple   & 2.155  & 13.10 & 0.409 & 5.20 & 39.69 & 0.690    & 13.24 & 0.90 & 6.80 & 0.413 & 12.88\\
		TM1e   & 1.957  & 12.45 & 0.496 & 5.11 & 41.04 & 0.790  &- & - & - & - & 13.02\\
        NL3   & 2.075  & 14.24 & 0.307 & 7.25 & 50.91 & 0.600    & 14.52 & 6.22 & 42.84 & 0.410 & 15.10 \\
		\hline\hline
		\end{tabular}}
		\label{tab:NSpropertiesS}
	\end{center}
\end{table*}

In Fig.~\ref{fig:2nbp}, we illustrate the EOSs for the hadronic phase, mixed phase, and quark phase using the NL3L-50, BigApple, and TM1e parameter sets, 
considering various strengths of vector couplings, $G_V=0,\ 0.1\ G_S,\ 0.2\ G_S$. 
Before a density of $\sim$0.7~fm$^{-3}$, the EOS lines for the hadronic phase using the NL3L-50 and BigApple parameter sets exhibit similarities, with BigApple being slightly softer. 
This results in a slight delay in the onset of the mixed phase, combined with an enlarged range of the mixed phase for all values of $G_V$. 
The relatively softer parameter set, TM1e, further enhances this trend. 
However, it is noteworthy that the mixed phase EOS derived from TM1e is stiffer compared to that of NL3L-50 and BigApple. 
This observation is counterintuitive since, despite the hadron-quark phase transition being delayed for a softer EOS (TM1e),
the stiffness of relatively soft EOS (TM1e) at the onset density of the mixed phase is still lower than that of the stiff EOS (NL3L-50 and BigApple). 
This is illustrated in Fig.~\ref{fig:3nbcs2}.
One plausible hypothesis is that a stiffer hadronic EOS results in higher energy within the mixed phase, 
causing nucleons to dissociate into quarks more rapidly than in a softer hadronic EOS.
Consequently, the rate of soft EOS stiffness change becomes smoother, as evident in Fig.~\ref{fig:3nbcs2}.
With increasing $G_V$,  the onset of the mixed phase shifts to higher densities, 
accompanied by an increase in the pressure of the mixed phase.
The degrees of freedom of the mixed phase encompass those of both the hadronic and quark phases,
contributing to the gentlest pressure change among the phases.
We utilize the squared sound velocity $c_s^2$ and the polytropic index $\gamma$ to quantify the stiffness of EOSs as illustrated in
Fig.~\ref{fig:2nbp}. 
The sound velocity $c_s$ is defined as 
$c_s^2={d {P}}/{d{\epsilon}}$, 
which asymptotically approaches $1/3$ in the conformal limit corresponding to free mass-less quarks. 
On the other hand, the polytropic index is defined as $\gamma={d({\mathrm{ln}P})}/{d({\mathrm{ln}\epsilon})}$, 
and has a value $\gamma=1$ in conformal limit matter.
The squared sound velocity (upper panel) and the polytropic index (lower panel) as functions of the baryon number density $n_b$ are depicted  in Fig.~\ref{fig:3nbcs2}. 
We show the results that $n_b$ stretched up to $2.0~\fm^{-3}$ only for comparison purpose 
where the hadronic matter models used in this study should not applicable at such high densities.
Both sound velocity and the polytropic index could characterize the stiffness change of the EOS, 
but sound velocity is the better choice in low densities because of it without a fluctuation.
The decrease of sound velocity of BigApple and TM1e may be related to the effect of parameter $\Lambda_{\rm{v}}$ in RMF models. 
The sudden decrease in the squared sound velocity $c_s^2$ or the polytropic index $\gamma$ at the onset of the mixed phase corresponds to an increase in degrees of freedom.
The values of $c_s^2$ and $\gamma$ for the mixed phase are the lowest among the three phases.
Towards the end of the mixed phase, the trend of $c_s^2$ approaches zero (especially for NL3L-50 and BigApple), 
resembling the results of the Maxwell construction.
$c_s^2\sim0$ corresponds to nearly constant pressure, as shown in Fig.~\ref{fig:2nbp}, 
the mixed phase pressure at high densities remains nearly constant for NL3L-50 and BigApple. 
This behavior arises because, 
at the end of the mixed phase in the Gibbs construction, 
the number density of leptons approaches zero, 
leading the hadronic and quark components to approach local charge neutrality, 
similar to the conditions of the Maxwell construction.
The value $\gamma=1.75$, from Ref.~\cite{Annala2020} distinguishes between the pure nucleon and non-nucleon parts:
all the hadronic matter EOS have $\gamma>1.75$ except the case TM1e-$G_V=0.2~G_S$, 
and all the mixed phase and quark phase EOSs have $\gamma<1.75$.

\begin{figure*}[!htbp]
\includegraphics[bb=10 30 540 480, scale=0.42]{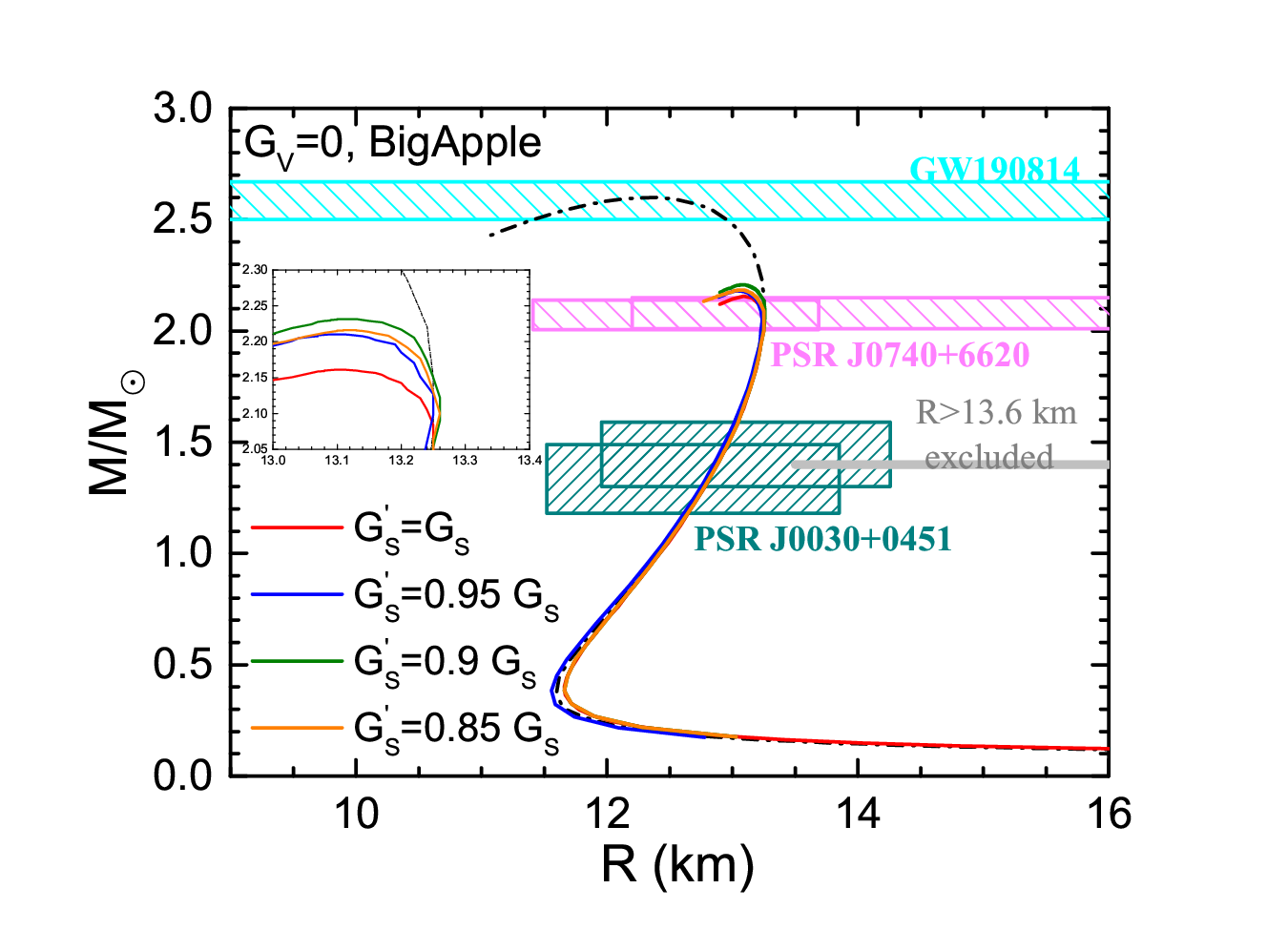}%
\includegraphics[bb=10 30 540 480, scale=0.42]{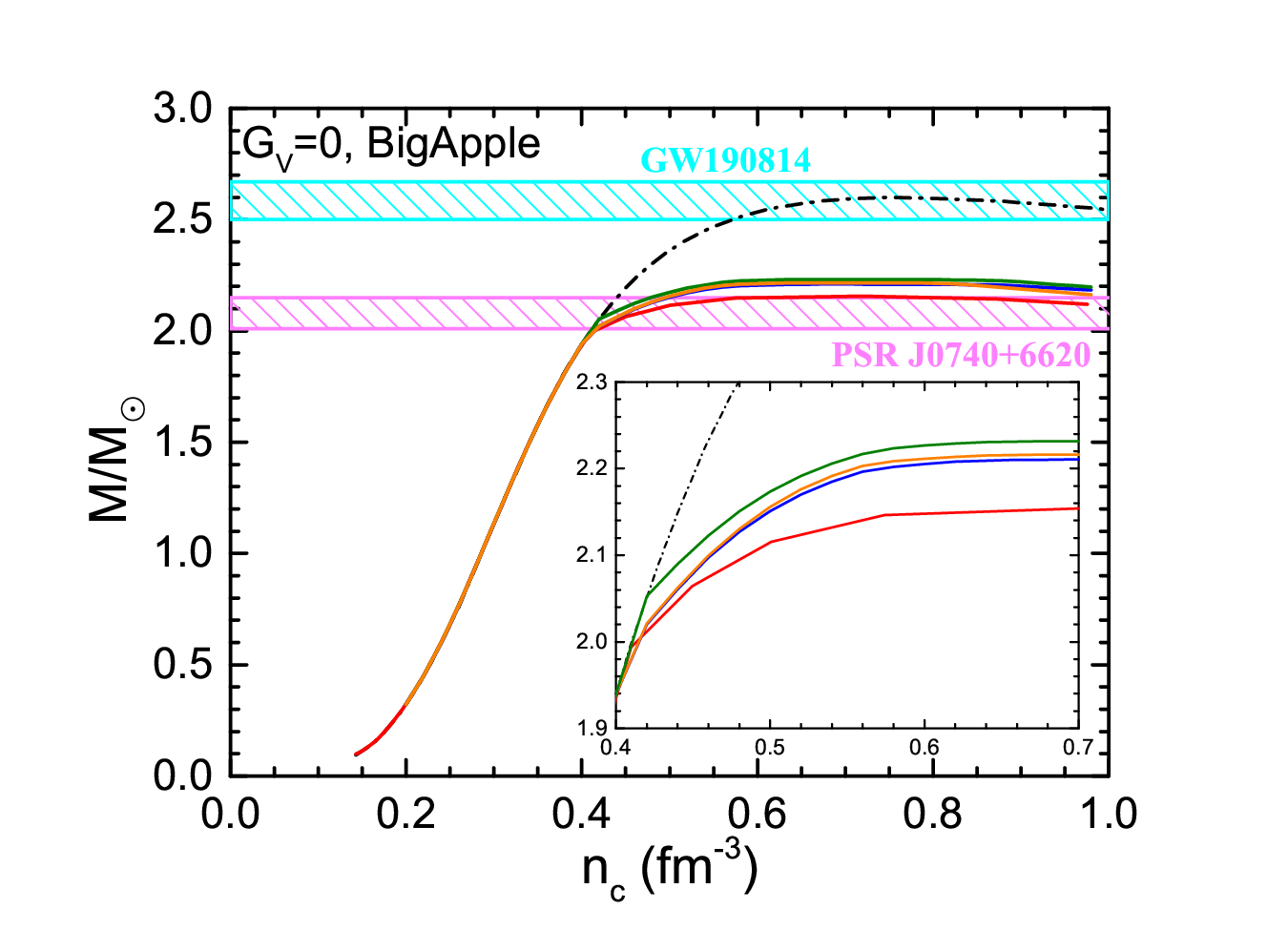}\\
\caption{Same as Fig.~\ref{fig:4rm} but for different values of $G_S^{'}$ with BigApple and $G_V=0$. 
The sub-figures show local enlargements.}
\label{fig:6gs}
\end{figure*}

\begin{figure*}[htb]
\includegraphics[bb=20 20 560 580, width=0.325\linewidth]{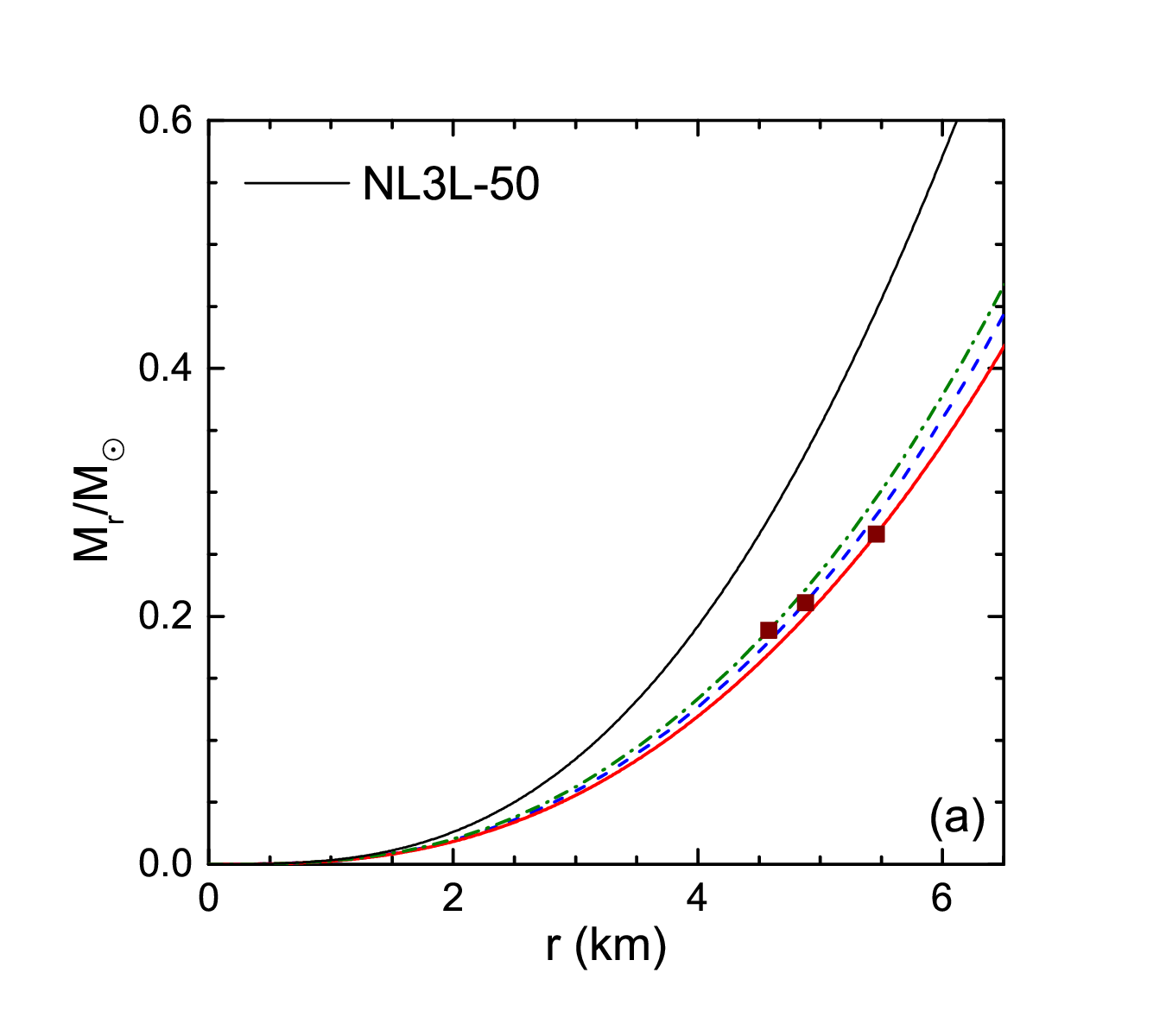}%
\includegraphics[bb=20 20 560 580, width=0.325\linewidth]{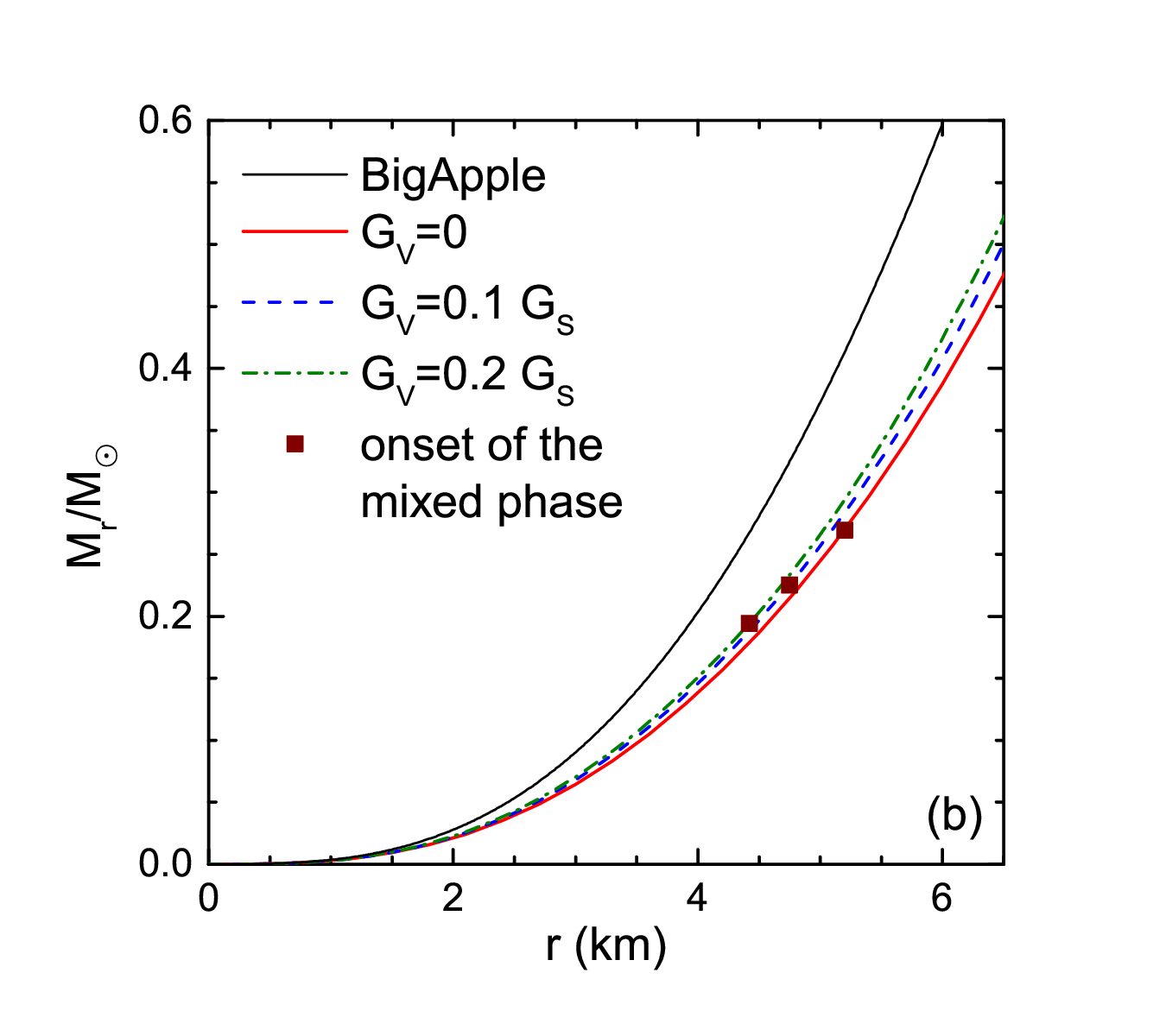}%
\includegraphics[bb=20 20 560 580, width=0.325\linewidth]{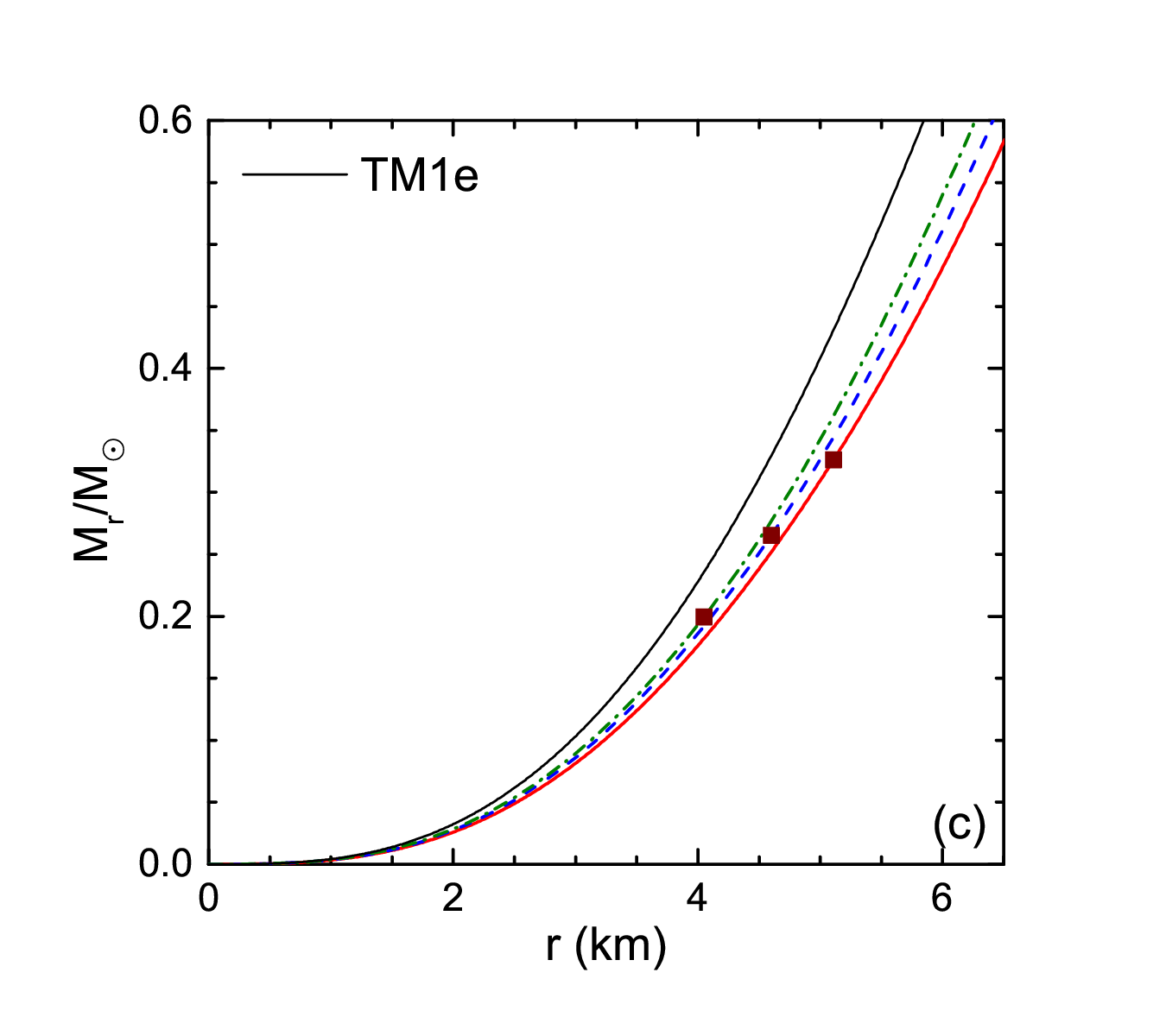}%
\caption{Mass internal to the radius $M_r$ changes with hybrid star internal radius $r$.
The comparison of pure hadronic matter in maximum-mass hybrid star results (black solid lines) are also shown. 
The solid square marks the onset of the mixed phase, with the left side indicating the mixed phase, while the right side representing the hadronic phase.}
\label{fig:7Mmaxrm}
\end{figure*}

In Fig.~\ref{fig:4rm}, we  display the mass-radius relations (left panel) 
and mass-central density relations (right panel) for NL3L-50, BigApple, and TM1e EOSs with different strength of vector coupling $G_V=0,\ 0.1\ G_S,\ 0.2\ G_S, 0.4~G_S$. 
Several constraints from astrophysical observations are also displayed in different color regions.
Several constraints from astrophysical observations are also displayed in different color regions.
The predicted maximum mass of hybrid star depends on $G_V$.
Due to the stiff enough EOS of hadronic matter, 
even with $G_V=0$, NL3L-50 and BigApple could support a maximum mass  of $M_\mathrm{max}\sim2.0~M_\odot$, 
while TM1e requires $G_V\geq0.1~G_S$.
The maximum mass could reach $2.5~M_\odot$ with $G_V\geq0.4~G_S$ using the NL3L-50 parameter set.
The appearance of quark degrees of freedom in the mixed phase leads to an obvious reduction of the maximum mass of hybrid star, 
but larger values of $G_V$ make the reduction of maximum mass smaller.
Compared with stiffer EOSs (NL3L-50 and BigApple), 
the effect of different values of $G_V$ on TM1e is smaller.
The normal $1.4~M_\odot$ NS properties are determined by the EOS around 2~$n_0$, where quarks do not yet appear.
Therefore, the mixed phase does not affect mass-radius line through PSR J0030+0451 constraints~\cite{Riley2019,Miller2019}.
Even when considering the hadron-quark phase transition, 
the radius and mass of hybrid stars are still mainly affected by the hadronic part of the EOS.
In Fig.~\ref{fig:5ml}, we show the dimensionless tidal deformability-mass relations for the NL3L-50, BigApple, and TM1e EOSs.
In the strong gravitational field generated by a neutron star's companion, the tidal deformability represents the deformation of a compact star and is related to its mass, radius, and Love number. 
From the binary neutron star merger event GW170817, the tidal deformability was extracted as $\Lambda_{1.4}=190^{+390}_{-120}$~\cite{Abbott2018}.
It can be found that, just like $R_{1.4}$, $\Lambda_{1.4}$ is also not affected by the hadron-quark phase transition.
Among these EOSs, the BigApple series fall within the constraints from GW170817, 
while the NL3L-50 and TM1e series slightly exceed this constraint.
To constrain the EOSs at high density over the phase transition point in this work, it is expected to measure the tidal deformability for massive neutron star mergers in the future.

To discuss the changes of the scalar coupling $G_S$ in NJL model, we display the same relations as shown in Fig.~\ref{fig:4rm}. 
The changing scalar coupling is introduced as $G_S^{'}=\frac{1}{\lambda^2}G_S$, where $G_S$ is the initial parameter with $G_S\Lambda^2=1.835$.
With $G_S^{'}<G_S$, the maximum mass of hybrid stars increases.
However, the maximum mass exhibit irregular variations rather than a monotonic change with the scalar coupling $G_S^{'}$.
Specifically, among the cases $G_S^{'}=0.85, 0.9, 0.95~G_S$ and $G_S^{'}=G_S$, the maximum mass is highest for $G_S^{'}=0.9~G_S$.
The mixed phase density range can be roughly identified in the right panel. 
The onset density is located at the split density with the pure nucleon star line and ends at the maximum mass position.
Indeed, for $G_S^{'}=0.85, 0.9, 0.95~G_S$ and $G_S^{'}=G_S$, the central densities of the maximum mass hybrid stars are around $n_c\sim0.7~\fm^{-3}$, similar to the results shown in Fig.~\ref{fig:1gvnb}(b).
An overall tendency shows a smaller mixed phase range compared to $G_S^{'}=G_S$.
The scalar coupling effect on quark matter in this work is different from that on color superconductivity (CSC) quark matter, as discussed in Ref.~\cite{Gholami2024b}.

\begin{figure}[htb]
\includegraphics[bb=40 5 580 580, width=8
cm,clip]{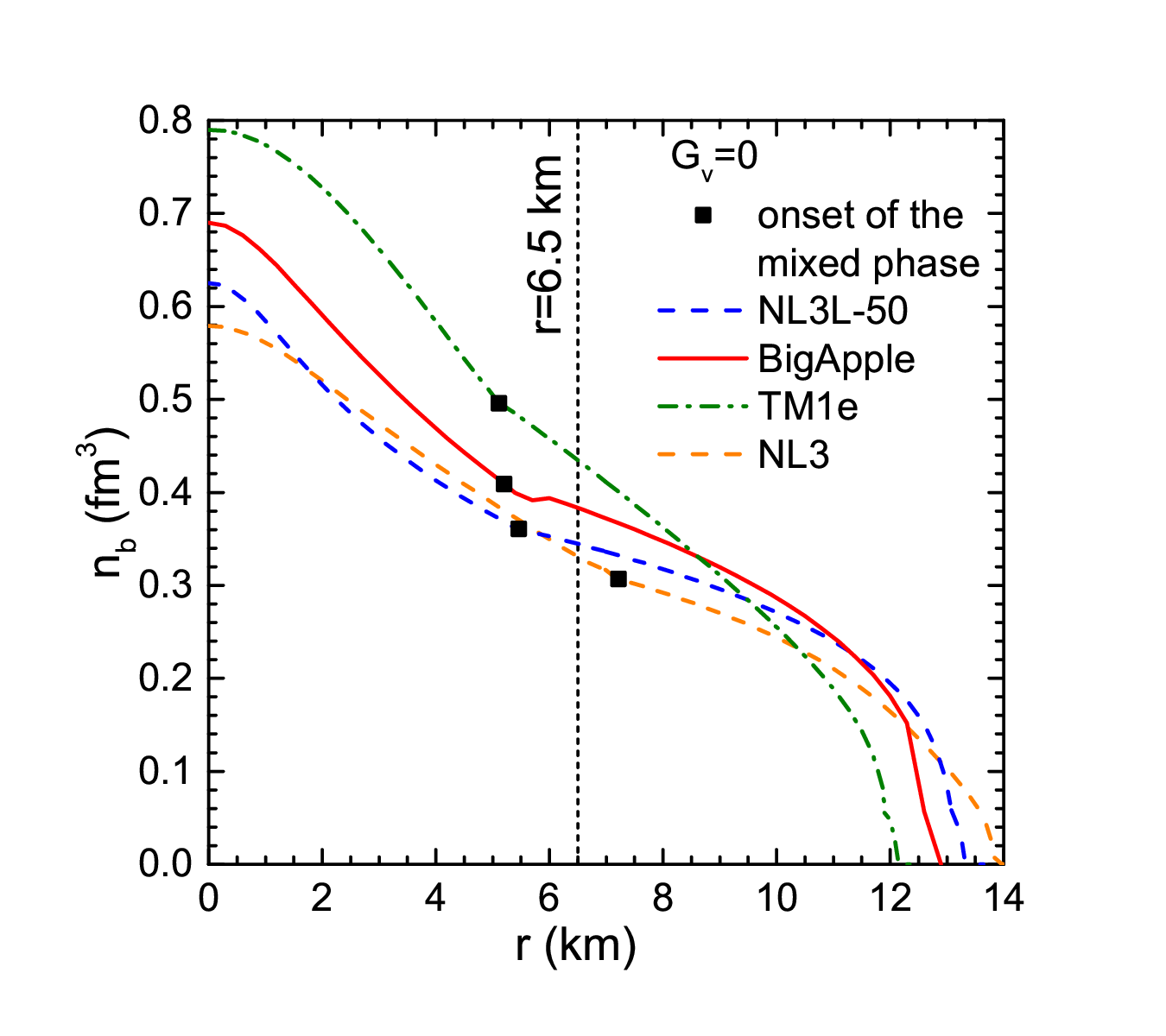}%
\caption{Internal density $n_b$ and internal radius $r$ relations of the maximum-mass hybrid star with $G_V=0$.
The line $r=6.5~\mathrm{km}$ represents approximately half the radius of the entire hybrid star.}
\label{fig:8rnb}
\end{figure}
\begin{figure}[htb]
\includegraphics[bb=40 5 600 580, width=8
cm,clip]{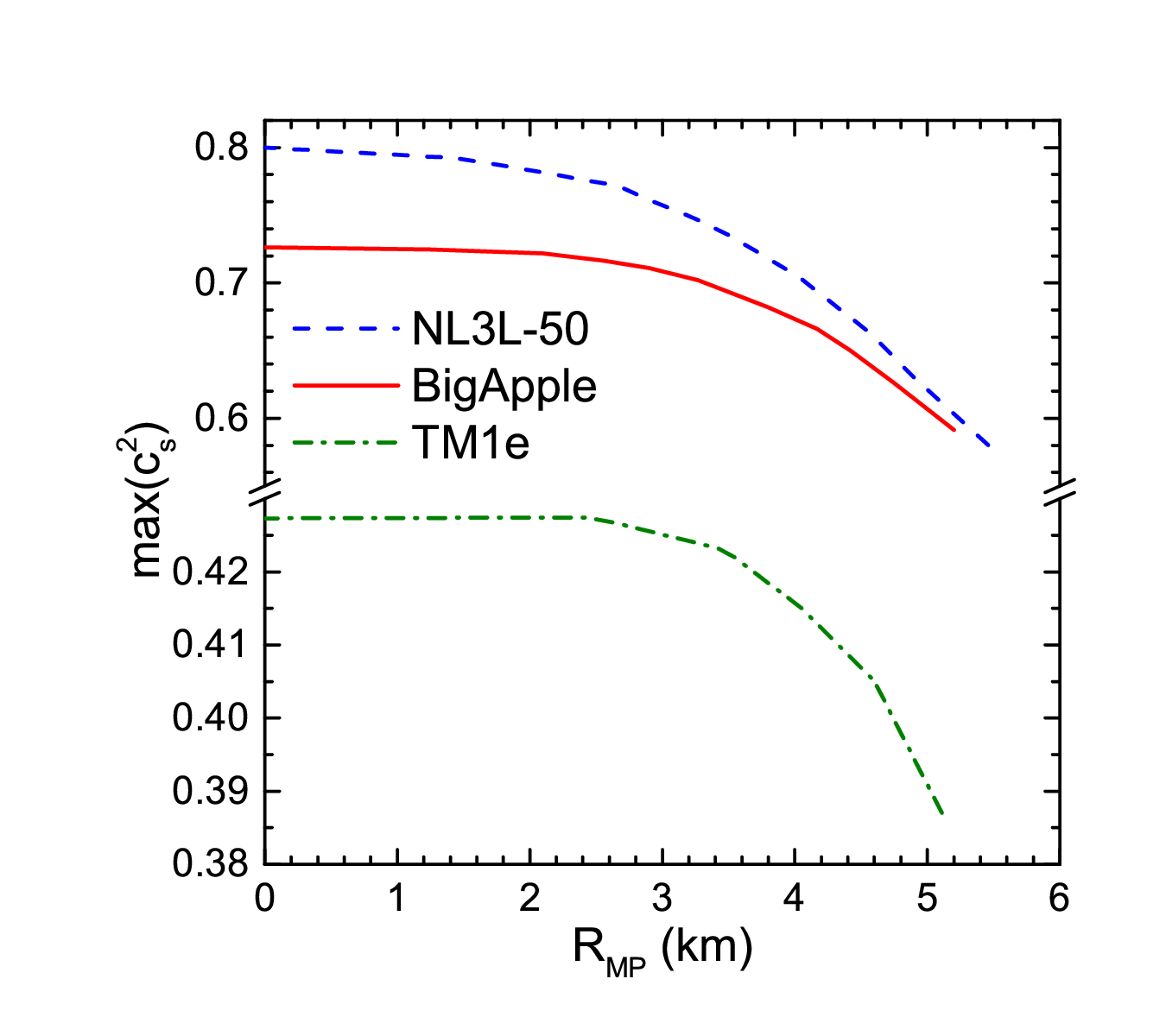}%
\caption{The relations between the radius of the mixed phase $R_{\mathrm{MP}}$  and the maximum squared sound velocity $\mathrm{max}~({c_s^2})$ of the maximum-mass hybrid star.}
\label{fig:9Rmpcs2}
\end{figure}

To better examine the properties of the maximum-mass hybrid star, 
we present the relations of internal mass $M_r$ and internal radius $r$ of the maximum-mass hybrid star in Fig.~\ref{fig:7Mmaxrm}.
For clarity, we use $M$ and $R$ to denote the total mass and radius of the hybrid star (like in Fig.~\ref{fig:4rm}), 
while $r$ and $M_r$ signify the internal radius within the hybrid star and the mass enclosed within that radius, respectively.
The results with mixed phased core (with $G_V=0,\ 0.1\ G_S,\ 0.2\ G_S$) and pure hadronic matter are shown.
The solid square marks the point at which deconfined quarks appear.
To the left of this solid square, the mixed phase core can reach sizes of approximately $6~\mathrm{km}$ 
and an enclosed mass of about $0.3~M_\odot$ for $G_V=0$.
As $G_V$ increases, the EOS for quark matter becomes stiffer,
resulting in a decrease in both the size and mass of the mixed phase core in the maximum-mass hybrid star.
A relative softer hadronic phase EOS (TM1e) could allow for a relatively large size and mass of the mixed phase core compared to stiffer hadronic phase EOS (NL3L-50 and BigApple), 
although it offers a relatively small total maximum mass.
No pure quark core is observed within the framework of this study, 
as the onset density for the pure quark phase exceeds the maximum central density, as shown in Fig.~\ref{fig:1gvnb}.
In contrast to results featuring a mixed-phase core, 
a neutron star composed solely of hadronic matter exhibits a rapid increase in internal mass $M_r$ as the internal radius $r$ expands. 
This increase in mass is evident throughout the entire interior of the star, 
rather than solely within the radius corresponding to the mixed-phase core.

Fig.~\ref{fig:8rnb} shows the internal density-radius ($n_b-r$) relations of the maximum-mass hybrid star,
including the NL3L-50, BigApple, TM1e, and NL3 parameter sets. 
Among these, the NL3 parameter set yields a larger mixed phase radius compared to the others
With $G_V=0$, NL3L-50, BigApple and TM1e could support a mixed phase core as large as $\sim40\%$ of total radius for the maximum-mass hybrid star.
In contrast, the NL3 model supports a mixed phase core where $R_{\rm MP}^{\rm max}/R_{\rm max}>1/2$ of the maximum-mass hybrid star,
however, it does not satisfy the radius constraints.
The bulk properties of maximum-mass hybrid stars and 2~$M_{\odot}$ hybrid stars with $G_V=0$ are summarized in Table.~\ref{tab:NSpropertiesS}.
Notably, the radius of the mixed phase in 2~$M_{\odot}$ hybrid stars is consistently smaller than that in maximum-mass hybrid stars.
This discrepancy arises because 2~$M_{\odot}$ stars have a lower central density, leading to a reduced range between  $n_b$(1) (the onset density of the mixed phase) and $n_c$.
Furthermore, although the NL3L-50 and BigApple models yield similar maximum masses with $G_V=0$, 
they predict significantly different sizes for the mixed phase core of a 2~$M_{\odot}$ hybrid stars ($R_{\rm MP}^{2.0}$).

By taking a series of values of $G_V$ like in Fig.~\ref{fig:1gvnb},
we obtain a series of phase transition positions and hybrid star maximum mass.
The relationship between the maximum value of squared speed of sound, $\mathrm{max}~(c_s^2)$, and the radius of the mixed phase of the maximum-mass hybrid star, $(R_\mathrm{MP})$, is exhibited in Fig.~\ref{fig:9Rmpcs2}. 
The $\mathrm{max}({c_s^2})$ is actually corresponds to the ${c_s^2}$ at density $n_b(1)$ for hadronic matter, $\mathrm{max}~({c_s^2})={c_s^2}~(n_b=n_b(1))$, as shown in Fig.~\ref{fig:3nbcs2}.
In the extreme case where $R_\mathrm{MP}=0$, indicating no mixed phase in neutron stars,
we have $\mathrm{max}(c_s^2)={c_s^2}~(n_b=n_c)$, 
under which condition the central density of the maximum-mass hybrid star is less than the onset density of the mixed phase.
Since $n_b(1)$ is influenced by $G_V$, within the same model,  
both $\mathrm{max}(c_s^2)$ and $n_b(1)$ increase with increasing $G_V$,
while $R_\mathrm{MP}$ decrease. 
This results in a smaller proportion of the mixed phase radius in the maximum-mass hybrid star.
This trend is consistent with the findings in Ref.~\cite{Annala2020}.

\section{Summary}
\label{sec:summary}
Different from less massive NSs, massive NSs may contain a quark core.
In this work, we focus on exploring the properties of maximum-mass hybrid stars with quark degrees of freedom, 
particularly the size of the quark matter core.
We investigate the EOS with the hadron-quark phase transition under Gibbs equilibrium. 
To achieve this, we employ the RMF model to characterize hadronic matter, 
while the NJL model with repulsive vector coupling is utilized to represent quark matter.
Our findings indicate that the properties of maximum-mass hybrid stars are significantly influenced by the model parameters, 
including the strength of the scalar coupling and vector coupling. 
The scalar coupling does not show a monotonic effect on the hybrid star maximum mass or the mixed phase range inside a hybrid star.
The mainly reason is that applying the RG consistency method leads to changes in the couplings and cutoffs as well.
Specifically, a stiffer hadronic matter EOS and a larger vector coupling are shown to potentially support more massive hybrid stars. 
The presence of vector coupling can increase the maximum mass of hybrid stars, 
partially mitigating the reduction in maximum mass caused by the emergence of quarks.
However, no clear correlation has been observed between the central density of maximum-mass hybrid stars and the vector coupling strength.
With the same vector coupling strength $G_V$,
a relatively softer hadronic matter EOS can support a larger mixed phase core, 
approximately 5 km in size.
This mixed phase core could occupy $\sim40\%$ of the entire neutron star when $G_V=0$, 
but the corresponding mass of the mixed phase core is comparatively small (less than 1/6) in relation to the total mass.
Our results suggest a significantly smaller mixed phase core in a $\sim2~M_\odot$ hybrid star compared to~\cite{Ferreira2021,Liu2023b,Annala2020}. 

In summary, our study reveals that the global properties of maximum-mass hybrid stars are sensitive to the strength of the maximum sound velocity of the EOS. 
Furthermore, it is possible for a sizable radius ($R_{\rm MP}^{\rm max}/R_{\rm max}\sim40\%$) of the mixed phase to exist in the core of maximum-mass hybrid stars with a small vector coupling.
However, for hybrid stars with $M=2~M_{\odot}$, 
only tiny mixed phase core is possible in this study.

\section*{Acknowledgment}
This work is supported by the National Natural Science
Foundation of China under Grants No. 12305148, No. 11975132, No. 12205158, 
Hebei Natural Science Foundation No. A2023203055,
and the Shandong Provincial Natural Science Foundation, China under Grants No. ZR2023QA112, No. ZR2022JQ04, No. ZR2021QA037, and No. ZR2019YQ01.


\newpage

\end{document}